\documentclass[fleqn,usenatbib]{mnras}

\usepackage[T1]{fontenc}
\DeclareRobustCommand{\VAN}[3]{#2}
\let\VANthebibliography\thebibliography
\def\thebibliography{\DeclareRobustCommand{\VAN}[3]{##3}\VANthebibliography}

\usepackage{graphicx}	
\usepackage{newtxtext}
\usepackage[varvw]{newtxmath} 
\usepackage{amsmath}	
\usepackage{xspace}     
\usepackage{tikz}
\usepackage{tikz-3dplot}
\usepackage{subfig}     
\usepackage{pdflscape}
\usepackage{standalone} 

\definecolor{darkpurple}{rgb}{0.6, 0.2, 0.8}
\definecolor{darkbrown}{rgb}{0.4, 0.26, 0.13}
\definecolor{darkblue}{rgb}{0.0, 0.0, 0.55}

\definecolor{grey}{rgb}{0.43, 0.5, 0.5}

\DeclareRobustCommand{\ion}[2]{%
\relax\ifmmode
\ifx\testbx\f@series
{\mathbf{#1\,\mathsc{#2}}}\else
{\mathrm{#1\,\mathsc{#2}}}\fi
\else\textup{#1\,{\mdseries\textsc{#2}}}%
\fi}

\newcommand{\HI}{{\ion{H}{I}}\xspace}
\newcommand{\HII}{{\ion{H}{II}}\xspace}
\newcommand{\Lya}{{Ly{$\alpha$}}\xspace}

\renewcommand{\d}[1]{\ensuremath{\operatorname{d}\!{#1}}}

\newcommand{\eg}{\mbox{e.g.\,}} 
\newcommand{\ie}{\mbox{i.e.\,}}

\newcommand{\eqn}[1]{Eqn.~(#1)}

\newcommand{\fig}[1]{Fig.~#1}
\newcommand{\Fig}[1]{Fig.~#1}

\title[Covariant 21-cm line radiative transfer]{A covariant formulation 
for cosmological radiative transfer \\ of the 21-cm line}

\author[Chan et al.]{
Jennifer Y. H. Chan$^{1,2,3}$\thanks{E-mail:jyhchan@cita.utoronto.ca (JYHC), 
qin.han.21@ucl.ac.uk (QH), kinwah.wu@ucl.ac.uk (KW), jason.mcewen@ucl.ac.uk (JDM)}, 
Qin Han$^{4}$, 
Kinwah Wu$^{4}$ and 
Jason D. McEwen$^{4}$ 
\\
$^1$Canadian Institute for Theoretical Astrophysics, University of Toronto, 60 St George St, Toronto, ON M5S 3H8, Canada \\ 
$^2$Dunlap Institute for Astronomy and Astrophysics, University of Toronto, 50 St George St, Toronto, ON M5S 3H4, Canada \\
$^3$Faculty of Arts and Science, University of Toronto, 100 St George St, Toronto, ON M5S 3G3, Canada \\
$^{4}$Mullard Space Science Laboratory, University College London, Holmbury St Mary, Surrey, RH5 6NT, UK
}

\date{Accepted 2024 April 19. Received 2024 April 2; in original form 2023 September 22}

\pubyear{2024}

\begin{document}
\label{firstpage}
\pagerange{\pageref{firstpage}--\pageref{lastpage}}
\maketitle

\begin{abstract} 

The 21-cm hyperfine line of neutral hydrogen is a useful tool to probe the conditions 
of the Universe during the Dark Ages, Cosmic Dawn, and the Epoch of Reionisation. In most of the current calculations, the 21-cm line signals at given frequencies are 
computed, using an integrated line-of-sight line opacity, with the correction for cosmological expansion. These calculations have not fully captured the line and continuum interactions in the radiative transfer, in response to evolution of the radiation field and the variations of thermal and dynamic properties of the line-of-sight medium. We construct a covariant formulation for the radiative transfer of the 21-cm line and derive the cosmological 21-cm line radiative transfer (C21LRT) equation. The formulation properly accounts for local emission and absorption processes and the interaction between the line and continuum when the radiation propagates across the expanding Universe to the present observer. Our C21LRT calculations show that methods simply summing the line optical depth could lead to error of $5\%$ in the 21-cm signals for redshift $z \sim 12-35$ and of $>10\%$ 
for redshift $z \lesssim 8$. Proper covariant radiative transfer is therefore necessary for producing correct theoretical templates for extracting information of the structural evolution of the Universe through the Epoch of Reionisation from the 21-cm tomographic data.
 
\end{abstract}  
\begin{keywords}
radiative transfer -- 
intergalactic medium -- 
dark ages, reionisation, first stars -- 
radio lines: general -- 
line: formation -- line profiles  
\end{keywords}


\section{Introduction}
\label{sec:introduction}

The Universe was once generally smooth and filled 
  with neutral gas \citep[][]{Smoot92, Bennett_13_9yrsWMAP, Planck13}, 
  mainly hydrogen, 
  but it looks very different today. 
The present Universe is structured, with
  gravitational-bound objects 
  forming the cosmic web, a giant network of brightly-lit gas, 
  stars and galaxies \citep{White1987ApJ_Nbody, Bond1996Nature, deLapparent1986ApJ, Colless2003astroph_2dFGRS,
  Tegmark2004ApJ_SLOAN, vandeWeygaert2009LNP_2dFGRS, Huchra2012ApJS_2MASSzSurvey, Guzzo2014AA_VIPERS, vandeWeygaert2016IAUS, Libeskind2018MNRAS, Bacon2021AA_MUSE}. %
The transition 
  that ushered the Universe 
  from suffusing 
  with neutral hydrogen (\HI) gas 
  into mostly ionised plasmas   
  -- the cosmological reionisation -- 
  was driven by 
  radiation from the first stars and galaxies, 
  and also the first Active Galactic Nuclei (AGN) 
  \citep[see \eg][]{BarkanaLoeb2001PhR, Pritchard2012Review, LoebFurlanetto2013Book, Mesinger2016Book, Dayal2020MNRAS, Munoz2022MNRAS}. 
The UV radiation 
  from the first stars and the first galaxies 
  carved the neutral medium surrounding them   
  into ionised cavities,  
  and the X-rays from the first AGN, 
  powered by accretion onto super-massive black holes, 
  easily turned a large amount of primordial atomic gases 
  into ionised plasmas over a large distance. 
As the Universe expanded and more ionising sources emerged, 
 these ionised cavities expanded,  
 percolated and merged 
 \citep[see \eg][]{Gnedin2000ApJ, Haiman20162003ApJ, Iliev2006MNRAS, McQuinn2007MNRAS, Adi2005MNRAS, Santos2008ApJ, Santos2010MNRAS,  Shin2008ApJ,  Robertson2010Nature, Mesinger2011MNRAS, Geil2017MNRAS, Kannan2022MNRAS}, 
  allowing ionising radiation to travel further  
  without further attenuation. 
The intergalactic space was gradually transformed,  
  from being a neutral atomic fog,    
  into a transparent sea of charged particles,   
  leaving only some islands of neutral gas 
  that are dense enough for self-shielding 
  from the ionising radiation 
  \citep[e.g.][]{Furlanetto2005MNRAS_TaxingTheRich, WyitheLoeb2008MNRAS}. 

How the development of structures in the Universe 
 was associated with the cosmological reionisation 
 is a fundamental question in astrophysics.  
The 21-cm hyperfine line of \HI 
  is identified as a means to 
  track the progression 
  of cosmological reionisation processes 
  and its interplay 
  with the formation of luminous objects, 
  i.e. stars, galaxies and accretion-powered compact objects \citep[see \eg][]{Field1959ApJ_reslineprofile, SunyaevZeldovich1975MNRAS_21cmGalaxiesBirth, HoganRees1979MNRAS_21cm, 
  Scott1990MNRAS, 
  Subramanian1993MNRAS, Madau1997ApJ,
  BarkanaLoeb2001PhR,
  Gnedin2004ApJ, 
  Zaldarriaga2004ApJ, 
  Furlanetto2006_review, 
  Morale2010ARAA,
  Pritchard2012Review, LoebFurlanetto2013Book, Furlanetto2016_21cmProbeEoR, Kulkarni2017MNRAS_21cmAGNEoR}.  
Observations to 
  map the 21-cm emission across the sky 
  over the evolutionary history of the Universe 
  have been proposed, and 
  some have already commenced, e.g. LOFAR\footnote{\url{https://www.astron.nl/telescopes/lofar/}},
  MWA\footnote{\url{https://www.mwatelescope.org/}}, 
  and HERA\footnote{\url{https://reionization.org/}}\citep[see e.g.][]{Koopmans2019ExA}. 
With the advent of the SKA\footnote{\url{https://www.skatelescope.org/}}, 
  we will be able to carry out 
  full-fledged all-sky 21-cm tomographic studies \citep[][]{Mellema2015aaska, Koopmans2015aska}. 
These all-sky data can also be used 
  for cross analysis with other data, 
  such as those obtained from survey observations \citep[e.g.][]{Furlanetto2007ApJ_21cmCrossCorrGalaxy} 
  and from line intensity mapping experiments  \citep[e.g.][]{Lidz2009ApJ, Visbal2010JCAP, Lidz2011ApJ_21cmCrossCorrCO, Carilli2011ApJ_CO, Gong2011ApJ_CO,  Silva2013ApJ_LyAlphaCrossCorr21cm, Chang2015aska_COCIILya, Breysse2022ApJ_COMAP, Silva2021ExA}.  
The reliability of 
  extracting information from the 21-cm tomographic data 
  of multi-wavelength cross-studies would depend 
  on our understanding of the production and attenuation 
  of the 21-cm line when it propagates 
  from the distant Universe to us. 
In this work we first present a covariant formulation 
  for the radiative transfer of the 21-cm line in an evolving and expanding Universe.  
The formulation takes proper account 
  of ionisation and thermal states 
  of the line-of-sight medium, 
  the evolution of the intergalactic radiation field,  
  and the cosmological expansion of the Universe. 
We then derive a ray-tracing scheme  
  for cosmological radiative transfer calculations 
  and compute the tomographic spectra 
  of the 21-cm line in representative settings. 

We organise the paper as follows. 
The covariant formulation of cosmological line radiative transfer is presented in Section~\ref{sec:CLRT}, followed by 
the construction of the radiative transfer equation for transporting the 21-cm line radiation in a flat expanding universe and 
the specification of the 21-cm line transfer coefficients in Section~\ref{sec:21cmCRT}. 
Section~\ref{sec:C21LRT-numerical} outlines the design of the cosmological 21-cm line radiative transfer (C21LRT) all-sky algorithm and its computational structure. 
The setting and input reionisation history for 21-cm tomography calculations with C21LRT in this paper are described in Section~\ref{sec:C21cmStudies}.
Section~\ref{sec:results} presents and discusses the 21-cm tomography results with C21LRT code, compares our results with the commonly adopted optical-depth parametrisation. The importance of proper modelling of line profiles, covariant formulation for C21LRT for correct tomographic studies is also explained. Finally, we summarise the paper in Section~\ref{sec:conclusion}. 

\section{Cosmological line radiative transfer}
\label{sec:CLRT}

\subsection{Line radiative transfer in the presence of continuum}
\label{subsec:LRT}

In a local rest frame, 
  the transfer equation for an unpolarised radiation of frequency $\nu$ and specific intensity $I_{\nu}$ reads 
 \begin{align}
\frac{\d I_{\nu}}{\d s}& = 
   -\left(\kappa_{{\rm C},\nu}+ 
  \kappa^{{\rm abs}}_{{\rm L},\nu}\, \phi_{\nu,{\rm abs}}  
   - \kappa^{{\rm sti}}_{{\rm L},\nu}\, \phi_{\nu,{\rm sti}} 
   \right) I_{\nu} \nonumber \\ 
   & \hspace*{1cm}
  + \left(\epsilon_{{\rm C},\nu}  
    +\epsilon_{{\rm L},\nu}\,\phi_{\nu,{\rm emi}} \right) 
\label{eq:generic-line-RT1}
\end{align} 
\citep[cf.][]{Wu2001MNRAS}, where $s$ is the photon's path length. 
The subscript ``C'' denotes the continuum underneath 
   and neighbouring to the line, 
  ``L'' denotes the line centre. 
  The absorption coefficient $\kappa_\nu$ has three components, 
  contributed by the absorption of the line and the continuum 
  and the stimulated emission
  (which can be considered as a negative absorption) of the line; 
  the emission coefficient $\epsilon_\nu$ has two components, 
  contributed by the emission of the line and the continuum. 
The line profile functions  
$\phi_{\nu,{\rm x}} \equiv 
\phi_{\rm x}(\nu -\nu_{{\rm line,0}})$,  
with ${\rm x} \in\{{\rm abs},\; {\rm emi},\; {\rm sti}\}$   
  corresponding to absorption, spontaneous emission 
  and stimulated emission, respectively,  
  are defined with respect to the 
   rest-frame frequency of the emission line, 
   and they are normalised, \ie  
\begin{align}
\int_{0}^{\infty} \d \nu\  \phi_{\nu,{\rm x}}
  & =  \int_{0}^{\infty} \d \nu\  \phi_{\rm x}
  (\nu -\nu_{{\rm line,0}})
  = 1 \  ,  
\label{eq:normalised} 
\end{align} 
   where $\nu_{\rm line,0}$
   is the intrinsic line  
   frequency, corresponding 
   the energy difference between two states 
   in the transition, 
   in the rest frame. 
This radiative transfer equation is valid 
  if the continuum at the line and its neighbouring frequencies 
  does not show strong variations, 
  e.g. no absorption edge. 
It is applicable 
  if photon scattering and 
  energy redistribution 
  between the electronic states are unimportant.  
This line profile function 
  gives the strength of the line,   
  in the context of absorption, spontaneous emission 
  and stimulated emission. 

Hereafter the notation, the subscript ``$\nu$'' 
 of the frequency-dependent quantities 
 are dropped, unless otherwise stated, 
 for better clarity in the expressions.    
Without losing generality, 
  suppose that the line has a symmetric profile 
  about its instrinsic line frequency, 
At the line centre, 
  the transfer equation (\eqn{\ref{eq:normalised}}) 
  takes the form 
\begin{align}
\frac{{\rm d} I_{\mathrm{L}}}{\mathrm{d} s} 
& = -\left(\kappa_{\mathrm{C}}+
\kappa^{{\rm abs}}_{{\rm L}}  
   - \kappa^{{\rm sti}}_{{\rm L}}
\right) I_{\mathrm{L}}
 +(\epsilon_{\mathrm{C}}+\epsilon_{\mathrm{L}}) 
\label{eq:lineRT_Wu01}
\end{align} 
\citep[see][]{Tucker1977Book, Wu2001MNRAS}. 
The transfer processes are determined by an opacity 
 contributed jointly by the line and the continuum. 
As the continuum is slowly varying with frequency,   
  the radiative transfer of the continuum 
  at the line frequencies 
  can be approximated 
  by the radiative transfer of the continuum 
  at the neighbouring frequencies  
  where the line profile function 
  is insignificant, \ie $\phi_\nu \ll 1$.  
This gives the continuum radiative transfer equation: 
\begin{align}
\frac{\d I_{\rm C}}{\d s}  & =  -\kappa_{\rm C} I_{{\rm C}}+\epsilon_{{\rm C}}  \   ,    
\label{eq:continuumRT_Wu01}
\end{align}   
  in which only the opacity of the continuum contributes to the transfer process. 

Whether the line appears as an emission feature or an absorption feature 
  depends on the relative strength of $I_{\rm L}$ and $I_{\rm C}$, 
  if the line is centrally peaked. 
The line appears as emission when $I_{\mathrm{L}} > I_{\mathrm{C}}$, 
  and in absorption when $I_{\mathrm{L}} < I_{\mathrm{C}}$.  

\subsection{Covariant formulation for generic line radiative transfer}\label{subsec:covarCLRT}

Radiative transfer in a cosmological setting  
  needs to firstly account for the expansion of the Universe.   
It needs to account also for the frequency redistribution 
  in the radiative processes 
  and the complex structures in the 
  frequency shifted and stretched line profile. 
The covariant form 
  of the radiative transfer equation  
  is given by 
\begin{align}
  \frac{\rm d}{{\rm d} \lambda_{\rm a}}  
   \left(\frac{I_\nu}{\nu^3} \right) 
  \;\!
 \Bigg\vert_{\lambda_{\rm a}, {\rm co}}  
 & = -k_{\alpha}u^{\alpha} \Big|_{\lambda_{\rm a},{\rm co}}  
  \left\{ -\kappa_{{\rm tot},\nu}  \;\!
   \left( \frac{I_{\nu}}{\nu^3} \right) 
   + \frac{\epsilon_{{\rm tot},\nu}}{\nu^3} \right\}    \ 
\label{eq:covarLRT123}  
\end{align} 
\citep[see][]{Younsi2012AA_GRRT, Chan2019MNRAS}, 
which is derived from the conservation of phase space density and the conservation of photon number. 
Here all the quantities are frequency dependent and are evaluated 
in a comoving reference frame (denoted by the subscript ``co'') 
 along the path of a photon. 
  If the processes that give rise to 
  the opacity of the line and the continuum are independent,
  the absorption and emission coefficients can be expressed as  
  the sum of the contributions of the relevant processes, i.e.\, 
$\kappa_{{\rm tot},\nu} = \kappa_{{\rm C},\nu} + {\tilde \kappa}_{{\rm L},\nu}$  and 
$\epsilon_{{\rm tot},\nu} = \epsilon_{{\rm C},\nu} + {\tilde  \epsilon}_{{\rm L},\nu}$, respectively. 
The factor $k_{\alpha}u^{\alpha}$ is arisen from evaluating the variation in the photon's path length $s$ with respect to the affine parameter $\lambda_{\rm a}$ for a photon with a 4-momentum $k^{\alpha}$ propagating in a cosmological medium that has 4-velocity $u^{\beta}$. 
In cosmological radiative transfer, 
 $k_{\alpha}u^{\alpha}$ 
 is determined by the space-time metric 
 of a chosen cosmological model. 
As such, \eqn{\ref{eq:covarLRT123}} 
 can be evaluated 
 in terms of a cosmological variable 
 instead of the affine parameter $\lambda_{\rm a}$ 
 \citep[see][]{Fuerst2004A&A}. 
In a flat Friedmann–Lema\^{i}tre-Robertson–Walker (FLRW) universe, 
  the metric has diagonal elements \,($-1 , a^2 , a^2 , a^2$), 
  where $a = 1/(1+z)$ is the cosmological scale factor and $z$ is the cosmological redshift.    
For a photon of 4-momentum $k^{\alpha}$ ($k^{\alpha}= (E/c, {\mathbfit{p}})$ where ${\mathbfit{p}} = (p_{r}, p_{\theta}, p_{\phi}) $ denotes the 3-velocity of the photon)  
  propagating in a cosmological medium with 4-velocity $u^{\beta}$ ($ u^{\beta}= \gamma (1, \,$ {\boldmath${\beta}$}) where {\boldmath${\beta}$} $ =(\beta_{r}, \beta_{\theta}, \beta_{\phi})$ denotes the 3-velocity of the medium and $\gamma = 1/\sqrt{(1+ \beta^2)}$ is the corresponding Lorentz factor) to the observer, 
\begin{align}
{k_{\alpha}u^{\alpha}\Big|_{z}} &= 
 \gamma_{z} \nu_{z} (-1+a^{2}\beta_{r, z})   \ ,
\end{align}  
where we have used the convention that the speed of light and the Planck constant have numeric value 1 (i.e., $c=h=1$) here.
  If the motion of the line-of-sight medium 
  is insignificant 
  (\ie $\beta$ = 0 and $\gamma$=1), 
the ratio of $k_{\alpha}u^{\alpha}$ evaluated at an earlier epoch to that at the present day is the relative shift of energy (or frequency) of the photon, i.e.  
\begin{align}
\frac{{k_{\alpha}u^{\alpha}\;\!\big|_{z}}}{k_{\beta}u^{\beta}\;\!\big|_{z_{\rm obs}}}
& = \frac{\nu_{\rm z}}{\nu_{z_{\rm obs}}}  
= \frac{a_{\rm obs}}{a}= \frac{1+z}{1+z_{\rm obs}}\    .
\label{eq:ratioCPRT}
\end{align} 

Note that in an expanding flat universe, 
 locally at a redshift $z$, we may express 
 the derivative by
\begin{align} 
\frac{\rm d}{{\rm d} \lambda_{\rm a}}\bigg\vert_z
  & =  
  \frac{{\rm d}\,x^{0}}{{\rm d}\lambda_{\rm a}}  
  \frac{\rm d}{{\rm d}\,x^{0}} \bigg\vert_z 
  = E \frac{\rm d}{{\rm d}s}  \bigg\vert_z 
  = \left( E\,\frac{{\rm d}z}{{\rm d}s}\right)\bigg\vert_z 
  \frac{\rm d}{{\rm d}z}  \   , 
\label{eq:chain}
\end{align} 
 where  
 $k^{\alpha} = (E/c, {\mathbfit{p}}) 
 = {\d {x^{\alpha}}}/{\d \lambda_{\rm a}}$.    
It follows, from Eqn.~(\ref{eq:covarLRT123}),   
  that the covariant line radiative transfer equation 
  for a flat FLRW Universe is  
\begin{align} 
\frac{\rm d}{{\rm d} z} \left(\frac{I_{\nu}}{\nu^3}\right)
  & = (1+z)\;\! \left[
    -\left(\kappa_{{\rm C},\nu}+ {\tilde \kappa}_{{\rm L},\nu}
    \right)  
      \;\!\left(\frac{I_{\nu}}{\nu^3}\right) 
   +  \frac{(\epsilon_{{\rm C}, \nu} 
     +{\tilde \epsilon}_{{\rm L},\nu}
  )}{\nu^3}\;\! 
     \right] 
     \frac{\d s}{\d z} \ . 
\label{eq:covariantlineRTstiemiInc}
\end{align}  
Here, 
  the line absorption coefficient ${\tilde \kappa}_{{\rm L},\nu}$ and the line emission coefficient ${\tilde \epsilon}_{{\rm L},\nu}$ account for the line profile, specified at a fixed redshift. 
These transfer coefficients are macroscopic variables,  
  but governed by microscopic processes. 
The increment of path length 
  with respect to the change in redshift is 
\begin{align} 
\frac{\d s}{\d z} 
  =\frac{c}{H_0}\,({1+z})^{-1}
  \big[\;\! \Omega_{{\rm r},0}(1+z)^4
  +\Omega_{{\rm m},0}(1+z)^3
   +\Omega_{\Lambda,0}\;\! \big]^{-1/2}  
\label{eq:dldz}
\end{align} 
\citep[see \eg][]{Peacock1999Book}, 
where $H_{0}$ is the Hubble parameter, 
$\Omega_{{\rm r},0}$, $\Omega_{{\rm m},0}$ and $\Omega_{\Lambda,0}$ are the 
dimensionless energy densities of relativistic matter and radiation, non-relativistic matter, 
and a cosmological constant (dark energy with an equation of state of $w \equiv -1$), respectively. The subscript ``0" here indicates that the quantities are measured at the present epoch (\ie $z=0$).  

\section{Cosmological 21-cm line radiative transfer}\label{sec:21cmCRT}

Here we present the cosmological transport specific to the 21-cm line of \HI in the presence of a background continuum radiation such as the cosmic microwave background (CMB). We also provide a pedagogical discussion about the global and local line shifting and line broadening effects that a line radiation would be subjected to in a general physical settings.

\subsection{The 21-cm line transfer equation and transfer coefficients}\label{subsec:lineTransCoef} 

The 21-cm line absorption and emission are determined by the probability of 
  the transition between the hyperfine states 
  (``u'' for the higher-energy triplet $1_1{\rm S}_{1/2}$ state,  
  and ``l'' for the lower-energy $1_0{\rm S}_{1/2}$ singlet state) 
  due to the spin-flip of the ground state electron 
  in the \HI. 
The transition between the two spin states 
is a magnetic dipole transition. 
It is mediated 
  by the emission or absorption of a photon 
   of an energy difference between the two hyperfine levels, 
   \ie $\Delta E_{\mathrm{ul}} = h \nu_{\rm ul}= 5.87 \times {10}^{-6}$~eV, 
   corresponding  to a wavelength of 21.1~cm
   and a frequency of $\nu_{21{\rm cm}} = 1.42~{\rm GHz}$ 
   \citep{Hellwig1970IEEETIM, Essen71hyperfine}.  
The line profile functions for 
   the \HI hyperfine line 
   are therefore 
   $\phi_{\rm x}(\nu -\nu_{\rm line,0})   
    = \phi_{\rm x}(\nu -\nu_{21{\rm cm}})$, 
    where ${\rm x} \in\{{\rm abs},\; {\rm emi},\; {\rm sti}\}$.

The transition probabilities 
  of the absorption, spontaneous emission and stimulated emission 
  of a photon with $\Delta E_{\rm ul}$  
  are specified by the Einstein coefficients    
  $B_{\rm lu}$, $A_{\rm ul}$ and $B_{\rm ul}$, respectively. 
Consider an ensemble of \HI atoms 
  in the $1{\rm S}$ ground state, 
  with the populations of electrons in its triplet and singlet spin states 
  specified by the number densities $n_{\rm u}$ and $n_{\rm l}$, respectively. 
Thus, the effective emission coefficient of the 21-cm line  
  may be expressed as 
 \begin{align}  
  \epsilon_{\rm{21cm}} & = \epsilon_{\rm{ul}}  
   = \frac{h\nu_{\rm{ul}}}{4\pi} \;
     n_{\rm{u}}A_{\rm{ul}} \int_0^\infty {\rm d}\nu\; \phi_{\nu, {\rm emi}} \ . \label{eq:21cmEmiCoewhole}  
\end{align} 
Similarly, the expression for the effective absorption coefficient 
  of the 21-cm line is  
\begin{align} 
  \kappa_{\rm L} & = \kappa^{\rm abs}_{\rm{21cm}}- \kappa^{\rm sti}_{\rm{21cm}} 
    \nonumber \\ 
   & = \kappa^{\rm abs}_{\rm{ul}}- \kappa^{\rm sti}_{\rm{ul}} \nonumber \\  
  & = \frac{h\nu_{\rm{ul}}}{4\pi} 
   \left[ \;\! n_{\rm{l}}B_{\rm{lu}}
      \int_0^\infty {\rm d}\nu\; \phi_{\nu,{\rm abs}} 
   - n_{\rm{u}}B_{\rm{ul}} \int_0^\infty {\rm d}\nu\; \phi_{\nu,{\rm sti}} 
   \;\!
      \right]  \ . 
   \label{eq:21cmAbsCoewhole}
\end{align}  
It follows 
  that the specific emission and absorption coefficients are 
\begin{align}  
  \epsilon_{\rm{L},\nu}  
    & = \frac{h\nu_{\rm{ul}}}{4\pi} \; 
      n_{\rm{u}} A_{\rm{ul}} \phi_{\nu, {\rm emi}} \ ; 
      \label{eq:21cmEmiCoefRT}   \\ 
  \kappa_{\rm{L},\nu} 
    & = \frac{h\nu_{\rm{ul}}}{4\pi} 
   \left[  n_{\rm{l}}B_{\rm{lu}}\;\!\phi_{\nu,{\rm abs}} 
         - n_{\rm{u}}B_{\rm{ul}}\;\!\phi_{\nu,{\rm sti}} 
      \right]      \  .
\label{eq:21cmAbsCoefRT}      
\end{align}

For a two-level system in thermal equilibrium, 
  characterised by a thermal temperature $T$, 
  the relative population of the particles at the two levels 
  differing by an energy $\Delta E_{\rm ba}$
  (with labels ``b'' and ``a'' for the levels 
  with the higher energy and lower energy, respectively), 
  is specified by the Boltzmann factor:  
\begin{align} 
  \frac{n_{\rm b}}{n_{\rm a}} & = \frac{g_{\rm b}}{g_{\rm a}}\; 
  {\rm exp} \left(-\frac{\Delta E_{\rm ba}}{k_{\rm B} T} \right)  \ ,   
\end{align} 
where $k_{\rm B}$ is the Boltzmann constant.
Analogous to the expression for the thermal system, 
  the relative population of the upper and lower hyperfine states 
  of the 21-cm line  
  may be expressed in terms of a temperature, $T_{\rm s}$, 
    known as the spin temperature \citep{Field1958IRE_WFeffects}: 
\begin{align} 
  \frac{1}{3} \left(\frac{n_{\rm u}}{n_{\rm l}} \right)   
   & =  {\rm exp} \left(-\frac{\Delta E_{\rm ul}}{k_{\rm B} T_{\rm s}} \right) 
    = {\rm exp} \left(-\frac{T_{\star}}{T_{\rm s}} \right) \  , 
\label{eq:Tspin}
\end{align}   
  with $g_{\rm u} = 3 g_{\rm l}$ (for \HI in the 1S ground state) 
  and $T_{\star} \equiv h\nu_{\rm 21cm}/k_{\rm B} 
  = \Delta E_{\rm ul}/k_{\rm B} = 0.0682~{\rm K}$. 
Note that when $T_{\mathrm{s}} \gg  T_{\star}$, 
  three of four atoms will be in the upper hyperfine level. 
However, there are mechanisms that can cause violation of this population partition. 
For example, \Lya pumping will allow 
  a higher number of \HI atoms 
  in the upper hyperfine state  
  through the Wouthuysen-Field mechanism  \citep{Wouthuysen1952AJ, Field1958IRE_WFeffects}.
Collisional excitation 
  is important in establishing 
  the population of electrons in the upper 
  hyperfine state in the absence of an external radiation field. The collision rate between particles generally increases with the square of particle number densities. Collisional excitation and de-excitation are, therefore, particularly important in 
  high-density environments.  
Collisions could be the dominant process 
  for the hyperfine transitions during the Dark Ages. However, they would give way to the radiative processes, when the first luminous objects began to appear. 
High-frequency radio background radiation at a high $z$ can be redshifted into the frequency of the 21-cm line, causing absorption or inducing stimulated emission. 
The background radio sources 
  can be the diffuse ambient CMB,  
  but can also be strong radio emitters, 
  such as quasars. 

Without loss of generality,  
  consider only the line absorption and  emission 
  and ignore the continuum and its opacity for the time being.  
Then, the radiative transfer equation is simply 
\begin{align}
\frac{\d I_{\nu}}{\d s}  & =  -  \kappa_{\rm{L},\nu} I_\nu +  \epsilon_{\rm{L},\nu}    
   =  -  \kappa_{\rm{L},\nu}\left( I_\nu -  S_{\rm{L},\nu} \right)    
    \label{eq:LRT_noContinuum}
\end{align} 
  in the local rest frame, 
  where $S_{{\rm L},\nu} = \epsilon_{{\rm L},\nu}/\kappa_{{\rm L},\nu}$ 
  is the source function of the line. 
(This can be justified 
  if the free-free processes, which 
  usually contribute to the continuum emission and absorption, 
  are insignificant, \ie  \ 
  $\kappa_{{\rm C},\nu} \ll \kappa_{{\rm L},\nu}$
  and $\epsilon_{{\rm C},\nu} \ll \epsilon_{{\rm L},\nu}$.)  
In terms of the Einstein coefficients, 
  the radiative transfer equation for the 21-cm line, therefore, is 
\begin{align} 
\frac{\d I_{\nu}}{\d s} 
 & =  - \frac{h\nu_{\mathrm{ul}}}{4\pi}\, 
   \Big[ \left( n_{\mathrm{l}} B_{\mathrm{lu}}\,\phi_{\nu, {\mathrm{abs}}}
   - n_{\mathrm{u}} B_{\mathrm{ul}}\,\phi_{\nu, {\mathrm{sti}}} 
    \right) \,I_{\nu} 
    -n_{\rm{u}} A_{\rm{ul}} \phi_{\nu, {\rm emi}}   \Big]   \ , 
\label{eq:LRT_noContinuum_Einstein}
\end{align}  
  where $\nu_{\rm ul} = \nu_{\rm 21cm}$, and the source function is 
\begin{align}
 {S_{{\rm L}, \nu}} 
 & = \frac{n_{\rm u} A_{\rm ul}\,\phi_{\nu,{\rm emi}}}   
    { n_{\rm l} B_{\rm lu}\,\phi_{\nu, {\rm abs}}
     -n_{\rm u} B_{\rm ul}\,\phi_{\nu, {\rm sti}}} \nonumber  \\ 
 & =\left( \frac{A_{\rm ul}}{B_{\rm ul}} \right)  
   \frac{(\phi_{\nu, {\rm emi}}/\phi_{\nu, {\rm abs}})}  
    {(n_{\rm l} B_{\rm lu}/n_{\rm u} B_{\rm ul} ) 
     -(\phi_{\nu,{\rm sti}}/\phi_{\nu,{\rm abs}}) }  \nonumber \\
 & =  \left(\frac{2 h{\nu_{\rm ul}}^3}{c^2}\right) 
  \frac{(\phi_{\nu,{\rm emi}}/\phi_{\nu,{\rm abs}})}  
    {(n_{\rm l}\;\! g_{\rm u}/ n_{\rm u}\;\! g_{\rm l}) 
     -(\phi_{\nu,{\rm sti}}/\phi_{\nu,{\rm abs}})}   \ . 
\label{eq:LineSourceFunc}
\end{align} 
The derivation of the source function here has not 
  assumed a thermal equilibrium. 
If local thermal equilibrium (LTE) is imposed, 
the source function will become the Planck function $B_{\nu}(T)$, recovering Kirchhoff's Law. It then follows that,
\begin{align*} 
\label{eq:approx_LTE}
\frac{n_{\rm u}}{n_{\rm l}} & =   
 \frac{g_{\rm u}}{g_{\rm l}}\,  
 {\rm exp}\left({-\frac{h\nu_{\rm ul}}{k_{\rm B}T}}\right)   
  \ ,      
\end{align*}  
 and the source function in \eqn{\ref{eq:LineSourceFunc}}  
  becomes 
\begin{align*}
    S_{{\rm L}, \nu}
    & = \left(\frac{2 h {\nu_{\rm ul}}^3}{c^2} \right) 
      \Bigg[ {\rm exp}
      \left(\frac{h \nu_{\rm ul}}{k_{\rm B}T}\right) -1 \Bigg]^{-1}   
       = B_{\nu}(T) \ .
\end{align*} 
The condition for a LTE between the radiation and the medium 
  is not always satisfied, especially when the transition 
  is coupled with an external radiative process. 
In this situation,   
\begin{align*} 
\frac{n_{\rm u}}{n_{\rm l}} & \neq    
 \frac{g_{\rm u}}{g_{\rm l}} \, 
 {\rm exp} \left({-\frac{h \nu_{\rm ul}}{k_{\rm B}T}}  \right) 
  \ ,      
\end{align*}   
  and the source function cannot be represented by the Planck function, 
   \ie the radiation is non-thermal, although the relevant radiative processes involved could be thermal processes themselves.

The line transfer equation can be further simplified using the relations between the Einstein coefficients in the absorption coefficient. 
With 
\begin{align}
\frac{B_{\rm lu}}{B_{\rm ul}} 
  = \frac{g_{\rm u}}{g_{\rm l}}  \ ,  
\quad\text{and}\quad 
\frac{A_{\rm  ul}}{B_{\rm ul}}  =
 \frac{2 h \nu^{3}}{c^{2}}\bigg\vert_{\nu = \nu_{\rm ul}}   
\end{align}
\citep{Einstein1916, Einstein1917},  
the absorption coefficient can be expressed as 
\begin{align}
\kappa_{{\rm L}, \nu}  
   & = \frac{h\nu_{\rm{ul}}}{4\pi}\, 
   \left(n_{\rm{l}} B_{\rm{lu}}\,\phi_{\nu, {\rm{abs}}} 
    - n_{\rm{u}} B_{\rm{ul}}\,\phi_{\nu, {\rm{sti}}}  \right) 
     \nonumber \\
   & = \kappa^{{\rm abs}}_{{\rm L}, \nu}\, \phi_{\nu,{\rm abs}}
   \left(1 - \Xi \right)  \ ,
\label{eq:21cmAbsCoef}
\end{align} 
where the normalised absorption coefficient is 
\begin{align}
\kappa^{{\rm abs}}_{{\rm L}, \nu} 
  = \frac{h \nu_{\rm ul}}{4\pi}
  \;\! n_{\rm l}B_{\rm lu} 
  =\frac{1}{8\pi}
  \left(\frac{c}{\nu_{\rm ul}} \right)^2 
  \left(\frac{g_{\rm u}}{g_{\rm l}}\right) 
\;\! n_{\rm l}A_{\rm ul}
  \  ,
  \label{eq:normalisedLineAbs}
\end{align} 
   and the factor for the stimulated emission is 
\begin{align}
\Xi  = \frac{n_{\rm{u}}}{n_{\rm{l}}}
   \frac{g_{\rm{l}}}{g_{\rm{u}}}\,
    \frac{\phi_{\nu,{\rm{sti}}}}{\phi_{\nu,{\rm{abs}}}}     \ .  
\label{eq:StiEmiFactor}
\end{align} 
The factor $\Xi$ can exceed unity  
  if the upper hyperfine level is sufficiently populated, 
   \ie when 
   $(n_{\rm u}/n_{\rm l})>(g_{\rm u}/g_{\rm l})= 3$  
   for $\phi_{\nu,{\rm sti}} = \phi_{\nu, {\rm abs}}$. 
This could occur during 
  the Epoch of Reionisation (EoR) 
  where a strong radiation field 
  can be created 
  by the first quasars, the first stars, 
  or the first galaxies.

It is now clear that the emission and absorption coefficients 
 of the radiative transfer of the hyperfine 21-cm line of \HI 
 can be computed from the Einstein coefficient $A_{\rm ul}$, 
 for spontaneous emission, 
 if the number density of \HI (which is $n_{\rm u} + n_{\rm l}$)  
 and the ratio of $n_{\rm u}/n_{\rm l}$ 
 (given by the spin temperature $T_{\rm s}$) 
 are known, 
 and the line profile functions $\phi_{\nu, {\rm x}}$,  
 with ${\rm x} \in\{{\rm abs},\; {\rm emi},\; {\rm sti}\}$    
 are specified.    

The line profile functions 
  depend only on the properties of the emitting and absorbing gas. 
Generally, there is no guarantee that 
  the line profile functions  
  $\phi_{\nu, {\rm{emi}}}$, 
  $\phi_{\nu, {\rm{abs}}}$, and $\phi_{\nu, {\rm{sti}}}$ 
  are the same.
However, the intrinsic width of the 21-cm hyperfine line 
  is insignificant 
  in comparison to the broadening of the line 
  due to other processes.  
In a local rest frame, 
  all atoms are subject to the same external line broadening processes. 
Thus the same line profile function is used for absorption, spontaneous emission and stimulated emission, i.e.
  $\phi_{\nu, {\rm{emi}}} = \phi_{\nu, {\rm{abs}}} 
  = \phi_{\nu, {\rm{sti}}}=\phi_\nu$, in the radiative transfer equation. 
  The continuum and line absorption and emission coefficients 
  are additive, 
  when there is no correlation between the continuum and the line opacities. 
Hence, the radiative transfer equation is 
\begin{align}
\frac{\d I_{\nu}}{\d s} & = 
-\left(\kappa_{{\rm C}, \nu} 
  +  \kappa^{\mathrm{abs}}_{{\rm L}, \nu}\, \phi_{\nu} \,[1-\Xi]\right)\;\!I_{\nu}
+ \left(\epsilon_{{\rm C},  \nu}+\epsilon_{{\rm L}, \nu}\right) \ ,   
\label{eq:lineRTstiemiInc2}
\end{align}   
where the contribution of stimulated emission ($1-\Xi$) is written out explicitly for clarity.

The covariant formulation of the cosmological transport of a generic line derived in \S~\ref{subsec:covarCLRT}, when applied to the 21-cm line of \HI, gives  
the C21LRT equation: 
\begin{align} 
\frac{\rm d}{{\rm d} z} \left(\frac{I_{\nu}}{\nu^3}\right)
  & = (1+z)\;\! \left[
    -\left(\kappa_{{\rm C},\nu}+ \kappa_{{\rm L}, \nu}\, 
    \right)  
      \;\!\left(\frac{I_{\nu}}{\nu^3}\right) 
       \right. + \left. 
   \frac{(\epsilon_{{\rm C}, \nu} 
     +\epsilon_{{\rm L}, \nu}) 
  }{\nu^3}\;\! 
     \right] 
     \frac{\d s}{\d z} \ ,
\label{eq:covariantlineRTstiemiInc_21cm}
\end{align}  
(cf.~\eqn{\ref{eq:covariantlineRTstiemiInc}}). 
Here, all the line transfer coefficients are as defined above (\eqn{\ref{eq:21cmEmiCoefRT}}, \eqn{\ref{eq:normalisedLineAbs}}, and \eqn{\ref{eq:StiEmiFactor}}), with the line profile functions substituted by the same $\phi_\nu$.

\subsection{Shaping the profile of the 21-cm line}\label{subsec:lineprofile}

\subsubsection{Line Shifting}
When an emitting medium and the observer are not 
  co-located and do not co-move,  
  the frequency of the emitted radiation 
  will appear to be shifted as seen by the observer.  
For the hyperfine 21-cm line 
  from distant astrophysical systems, 
  the shift of the line can be caused by 
  (i) cosmological expansion, 
  and (ii) the relative velocity between the sources 
    and the observer. 

The former is a global effect. 
It leads to a redshift of the frequency of the radiation, 
  and it can be manifested in the shift 
  of centre frequency of the hyperfine 21-cm line 
  to a lower frequency at the observer's reference frame.  
Quantitatively, 
  the relative frequency shift at two cosmological locations or epochs 
  $z$ and $z'$ is given by  
\begin{align} 
\frac{\nu(z')}{1+z'} & = \frac{\nu(z)}{1+z} \ . 
\label{eq:cosmologicalFreqShift}
\end{align}  
In the context of radiative transfer and spectral evolution, 
  the frequency redshift of radiation 
  caused by cosmological expansion 
  is essential to a flow of photons from the high frequencies 
  to the low frequencies 
  at a constant rate 
  if evaluated in terms of the cosmological redshift $z$. 

The latter is associated with the local movement 
  of the emitter with respect to the observer. 
The frequency shift is simply a Doppler effect.  
In the non-relativistic limit, 
  the frequency of the centre of the observed 21-cm line $\nu'$ is 
\begin{align}
  {\nu'}_{\rm 21cm} = \nu_{\rm 21cm} \;\!
  (1 \pm v_{\parallel}/c) \  ,
\label{eq:DopplerShiftedfreq}   
\end{align} 
   where $v_\parallel$ 
   is the relative line-of-sight velocity of the \HI gas 
   with respect to the observer.  
$v_\parallel$ can be positive or negative.  
The 21-cm line will be shifted to higher frequencies  
  if the emitting \HI gas is approaching, 
  and to lower frequencies if the emitting \HI gas is receding with respect to the observer. 

\subsubsection{Line broadening and line profile}
\label{subsubsec:LineProfile}
The 21-cm hyperfine line 
is broadened by radiative damping, particle collision, 
   thermal motion and turbulence. 
The first two are intrinsic  
  to individual emitters,   
  where the line emission processes 
  are effectively damped oscillations.  
The last two are 
  manifestations of Doppler effect, 
  caused by the incoherent movements 
  of a collection of \HI gas particles  
  with respect to the observer. 

The damped oscillations associated with the emission process   
 will lead to the broadening of the 21-cm hyperfine line, 
 resulting in a Lorentzian profile 
  \citep[see \eg][]{Rutten2003RTNotes}: 
\begin{align} 
   \phi\left(\nu-\nu_{\rm 21cm} \right) 
   & =  \frac{1}{\pi} 
  \left[\;\! \frac{\left(\Gamma_{\rm all}/4\pi\right)} 
  {\left(\nu-\nu_{\rm 21cm} \right)^2
  +\left(\Gamma_{\rm all}/4 \pi\right)^2}\;\! \right]  \ ,   
  \label{eq:Lorentzian}
\end{align} 
 where $\Gamma_{\rm all}$ 
  is the sum of the damping parameters 
  (\ie reciprocals of the damping timescales) 
  of all the uncorrelated damping processes. 
Spontaneous emission and collision-induced emission 
  are independent processes, 
  and hence, 
  their combined broadening will be specified 
  by a total damping parameter 
\begin{align} 
  \Gamma_{\rm all}&
  = \Gamma_{\rm rad} + \Gamma_{\rm coll} 
  \propto\left( \frac{1}{t_{\rm rad,spon}} 
   + \frac{1}{t_{\rm rad,coll}}  \right) \ , 
   \label{eq:damping-para}
\end{align}  
  where $t_{\rm rad,spon}$ ($=1/A_{\rm ul}$) 
  and $t_{\rm rad,coll}$ 
  are the timescales 
  for the spontaneous emission 
  and for the collision-induced emission, 
  respectively. 
 During the EoR, 
  the Universe was sufficiently dense such that 
  the timescale of collisional de-excitation was significantly 
  shorter than the timescale for spontaneous emission\footnote{
The collisional timescale 
    is inversely proportional 
    to the density of the gas. 
The relative importance  
  of collisional damping and radiative damping 
    is, therefore, environment dependent
     in the 21-cm hyperfine transition.  
While collisional damping 
    dominated in 
    \HI gas suffusing the early Universe, 
  it is less important  
    in the present-day intergalactic medium.},   
   implying that $\Gamma_{\rm coll} \gg \Gamma_{\rm rad}$.  
The total broadening of the two processes combined 
  is, therefore, 
  $\Gamma_{\rm all} \approx \Gamma_{\rm coll}$, 
  corresponding to a full-width-half-maximum (FWHM) in frequency, ${\rm FWHM}_{\nu}$, of $(\Gamma_{\rm coll}/2\pi)$  
  in the Lorentzian line profile.  
   
The thermal motion and turbulent motion  
  of the gas particles 
  will give rise to Doppler shifts 
  in the radiation that they emit. 
The incoherent Doppler shifts of the 21-cm line emitted    
  from an ensemble of \HI gas particles  
  with different line-of-sight velocities 
  would make the line appear to be broadened. 
When the \HI gas particles 
  have a Gaussian velocity distribution,  
  the line will have a Gaussian profile: 
\begin{align}
\phi(\nu - \nu_{\rm 21cm}) 
  &  =\frac{1}{\sqrt{\pi} \;\! \Delta \nu_{\rm D}} \ 
{\rm exp} \left[-\left(
\frac{\nu-\nu_{\rm 21cm}}{\Delta \nu_{\rm D}}\right)^{2}\right] \  .  
\label{eq:Gaussian}
\end{align}
Suppose that the turbulent motion in a \HI gas 
  has a well-defined characteristic mean-square 
  velocity, $v^2_{\rm turb}$.  
Then, a velocity dispersion can be assigned for the \HI gas particles,  
  analogous to the thermal velocity dispersion, 
  and from it, a Maxwellian velocity distribution can be constructed.  
As thermal motion and turbulent motion are independent, 
  their velocity dispersions are additive. 
The effective width of the broadened line is  
\begin{align}
\Delta \nu_{\rm D} & \equiv  
 {\nu_{\rm 21cm}} \left(\frac{b_{\rm D}}{c}\right) 
  = \nu_{\rm 21cm}  
   \sqrt{\frac{2 k_{\rm B} T_{\rm k}}{m c^2} 
     + \left(\frac{v_{\rm turb}}{c}\right)^2} \  ,     
\label{eq:effdopplerwidth}
\end{align} 
    where  
\begin{align}
b_{\rm D} & = \sqrt{\frac{2 k_{\rm B} T_{\rm k}}{m} 
 + {v_{\rm turb}}^2}    \   
\label{eq:dopplerpara} 
\end{align} 
  is the Doppler parameter.  
For a Gaussian profile function, 
   the line has a 
  ${\rm FWHM}_{v} = 2\sqrt{\ln{2}} \ b_{\rm D}$ 
  in velocity. 
In frequency, 
\begin{align} 
  {\rm FWHM}_{\nu} 
   & = 2\sqrt{\ln{2}}\;\!  (\nu_{\rm 21cm})    
   \left(\frac{b_{\rm D}}{c} \right)    \nonumber \\ 
    & =   2\sqrt{\ln{2}}\;\! (\nu_{\rm 21cm}) 
     \sqrt{\frac{2 k_{\rm B} T_{\rm k}}{m c^2} 
     + \left(\frac{v_{\rm turb}}{c}\right)^2}  \ . 
\label{eq:fwhm:thermal_turbulence}
\end{align}  

The damping-induced line broadening is associated 
  with the internal action and response of the emitters;  
  the velocity-induced line broadening is associated 
  with the kinetics of the emitters. 
These two broadening are different by nature. 
Their effects are not additive, 
  despite that damping and velocity-induced broadening 
  are independent.  
  The total broadening 
  will be determined by a convolution 
  when both broadening processes are present. 
The convolution of a Lorentzian line profile 
  and a Gaussian line profile 
  is a Voigt line profile.  
Although Voigt profiles  
  do not have simple analytic form 
  in terms of elementary functions   
  \citep[see \eg][]{Schreier1992VoigtComputation, Boyer14VoigtEval, Mohankumar2019VoigtComputation, AlOmar2020VoigtProfile}, 
  the normalised Voigt profile for the 21-cm line 
  can be expressed as an implicit function of frequency 
  $\nu$ specified by three parameters: the line frequency centre, $\nu_{\rm 21cm}$,   
  the damping parameter, $\Gamma_{\rm all}$,  
  and velocity-induced width, $\Delta \nu_{\rm D}$ 
\citep[see][]{Rutten2003RTNotes, Chan2020Thesis}. 
In the astrophysical environments 
during cosmological reionisation,  
damping-induced broadening is unimportant 
when compared with velocity-induced broadening\footnote{Velocity-induced broadening is predominant in high-temperature, low-pressure environments where the thermal motion of particles is significant. Damping-induced broadening becomes more comparable or dominant in dense, high-pressure environments with frequent particle collisions but the temperature affecting the velocity of the particles must not be too high. In the context of cosmological reionisation eras probed by the 21-cm line, the low density of the IGM makes velocity-induced broadening the primary mechanism. While damping-induced broadening may increase in regions with higher densities due to structure formation, rising gas temperatures due to radiation from these first luminous structures further enhance velocity broadening. Thus, our radiative transfer calculations primarily focus on velocity-induced broadening.}. 
Thus, a Gaussian line profile is adopted 
in our subsequent radiative transfer calculations. By doing this, we let the emission or absorption distributes over an the effective frequency range $\sim$($\nu_{\rm 21cm}$-FWHM$_{\nu}$,$\nu_{\rm 21cm}$+FWHM$_{\nu}$). Then the emission and coefficients at line centre  $\kappa_{\rm L, \nu21cm}$ and $\epsilon_{\rm L, \nu21cm}$ are inversely proportional to FWHM$_{\nu}$. The emission or absorption integrated over the full relevant frequency range for a given line profile is conserved.

\subsection{Continuum radiation background}\label{subsec:ConRadBg}

Determination of the tomographic 21-cm line signal 
  requires simultaneously and self-consistently 
  solving the radiative transfer of the line 
  (\ie \eqn{\ref{eq:covariantlineRTstiemiInc_21cm}}) 
  and the radiative transfer of the continuum: 
\begin{align} 
\frac{\rm d}{{\rm d} z} \left(\frac{I_{\nu}}{\nu^3}\right)
  & = (1+z)\;\! \left[
    -\kappa_{{\rm C},\nu}
      \;\!\left(\frac{I_{\nu}}{\nu^3}\right) 
  + \frac{\epsilon_{{\rm C}, \nu}}{\nu^3}\;\! 
     \right] 
     \frac{\d s}{\d z} \ . 
\label{eq:covariantConRT}
\end{align}  
The omnipresence of the CMB photons provide the radiation background that must be accounted for when looking at the cosmological 21-cm line. 
The CMB's spectra at different redshifts are well described by the Planck function at its characteristic temperature $T_{\rm CMB}(z) = T_{\rm CMB, 0}(1+z)$, with $T_{\rm CMB, 0} = 2.73$~K \citep[][]{WilsonPenzias1965ApJ, Mather1994ApJ, Spergel2003ApJS, Planck13}. 

In addition to the CMB, 
diffuse continuum radiation can arise from local free-free processes and synchrotron radiation. 
The continuum absorption ($\kappa_{{\rm C},\nu}$) and emission ($\epsilon_{{\rm C}, \nu}$) coefficients for the thermal free-free process and the non-thermal synchrotron radiation appropriate for the studies of cosmic plasmas are discussed in Section 2.4 of \citet{Chan2019MNRAS}, with their explicit expressions presented in Appendix~C of that paper and references therein. 
The absorption and emission in the continuum at frequencies of the line and adjacent to the line are less relevant, although electron scattering 
could cause a certain degree of extinction 
when the line photons traverse the ionised matter along the line-of-sight. 
To isolate the 21-cm line emission and absorption, which is the focus of this work, the continuum emission and absorption are assumed to be zero. 

In the presence of a continuum background 
  of a bright source, 
  \eg a radio-loud quasar whose emission 
  is generally non-thermal synchrotron radiation 
  from relativistic electrons 
  and it tends to have a very high brightness temperature \citep[\eg][]{Willott1998RadioLoudQSOs,  Vernstrom2018QuasarRadio}, 
 the 21-cm line will appear as absorption. 
  Distant quasars 
     have been identified as candidate sources 
     for the detection of 21-cm forests \citep[][]{Carilli_02_Forest21cm, FurlanettoLoeb_02_Forest21cm,  Furlanetto_06_Forest21cm, Xu_09_Forest21IGMTemp, Mack_12_Forest21cmDetect, Ciardi_15_Forest21SKA}, where photons emitted at frequencies $\nu > \nu_{\rm ul}$ 
  by the bright background radio quasar 
  at redshift $z_{\rm emi}$ 
   are absorbed by the diffuse \HI gas 
   along the line-of-sight 
   at redshift $z = [\nu_{\rm ul}(1+z_{\rm emi})/\nu -1]$. 
Other candidate bright point sources 
  would be hypernovae, which show gamma-ray burst, with radio afterglows, and some could be fast radio burst (FRB) sources. 
However, little is known about 
  the number distribution of hypernovae and FRBs 
  and their evolution at very high redshifts, 
  such as $z\sim 6$ or higher.    

\subsection{Differential brightness}\label{subsec:diffTb}

The 21-cm line is observed against a continuum background, which is generally sourced from the CMB, and occasionally from bright radio sources. 
The observed differential brightness temperature is 
\begin{align}
\delta T_{\rm b} 
= (I_{{\rm L},\nu}-I_{{\rm C},\nu})
\frac{({{c}/{\nu})^2}}{2k_{\rm B}} \ , 
\label{eqn:diffBrightness}
\end{align}
where we have used the Rayleigh-limit approximation. 
In a self-consistent manner, 
$I_{{\rm L},\nu}$ is accurately determined by  \eqn{\ref{eq:covariantlineRTstiemiInc_21cm}} while  $I_{{\rm C},\nu}$ by \eqn{\ref{eq:covariantConRT}}.

\section{Numerical Radiative Transfer}
\label{sec:C21LRT-numerical}

\subsection{Ray-Tracing Scheme}
\label{subsec:C21LRT-raytrace}

The C21LRT 
 calculations consist of three key elements:  
(i) a ray-tracing algorithm accounting 
  for the transport of radiation from the past to the observer  
  in an expanding Universe, 
(ii) a computational component to determine 
  the interaction between the incoming background radiation and the local medium 
  and to evaluate the absorption of the incoming background radiation 
  and emission in the local medium, and  
(iii) a numerical solver for the C21LRT equation  
  (\eqn{\ref{eq:covariantlineRTstiemiInc_21cm}}) and the transfer equation of the continuum (\eqn{\ref{eq:covariantConRT}}) 
  along with the ray-tracing calculations. 

On cosmological scales, the transfer of radiation along a line of sight is the transfer of radiation from the past to the present. 
C21LRT calculations for a single ray will provide a tomographic 21-cm spectrum at the observer epoch $z=0$ (or at any arbitrary redshift from the line emission epoch). 
Pencil-beam calculations can also be performed wherein a bundle of rays labelled by their spatial markers are traced over their path (stamped by comic time) as they traverse through an extended medium, producing a map that represents a collection of end-points of the path-integrated rays. 
Each of these end-points is associated with its own tomographic 21-cm spectrum that would be observed today at $z=0$. 
In an all-sky setting, each ray can be marked by $(\theta, \phi)$ that corresponds to its celestial sky coordinates and the radial axis $r$ that corresponds to the redshift axis $z$. 
The C21LRT equation is then solved in a spherical polar coordinate system $(r, \theta, \phi)$, 
  as in the all-sky polarised radiative transfer 
  calculations of \citet{Chan2019MNRAS}. 

In the covariant formulation constructed for the C21LRT calculations, 
  the emission and absorption processes 
  are evaluated in a local rest frame. 
This allows 
  the physical variables,  
  and hence, the emission and absorption coefficients 
  and their changes along a ray, 
  to be parametrised by the cosmological redshift only.  
The propagation of the radiation is parametrised by the redshift $z$  
 which is divided into $N_{z}$ discrete cells along each ray. 
In addition to the discretisation along the ray, 
  another sampling in frequency 
  on each $z$-grid is constructed  
  to account for the changes 
  in the line profile and the continuum along the ray. 
The sampling scheme has been constructed to optimise the efficiency in the line radiative transfer calculations. 
The design of the algorithm and its optimisation are described in Appendix~\ref{app:C21LRTAlgo}.

Solving the cosmological 21-cm line radiative transfer equation, 
  \eqn{\ref{eq:covariantlineRTstiemiInc_21cm}},   
  requires 
  (i) the specification of initial radiation background, and 
  (ii) the determination 
  of the continuum and the line transfer coefficients 
  over a frequency range 
  fully covering the line and the relevant underlying continuum in the local rest frame. 
For (i), the CMB provides the continuum radiation background that must be accounted for when considering the cosmological 21-cm line. If bright radio sources are present along some rays, their emission at and adjacent to the 21-cm line centre will be the dominant background radiation. 
As per (ii), the line transfer coefficients (\eqn{\ref{eq:21cmEmiCoefRT}} and \eqn{\ref{eq:21cmAbsCoefRT}}) 
  can be computed 
  when the line profile function $\phi_{\nu}$ 
  and the number densities of \HI atoms $n_{\rm HI}$ 
  in the two hyperfine states $n_{\rm{l}}$ and $n_{\rm{u}}$ 
   at each location along the ray are known. 
Modelling of either $n_{\rm{l}}$ or $n_{\rm{u}}$ and $n_{\rm HI}$ in astrophysical environments involves detailed investigations of the spin-flip mechanisms at play and is beyond the scopes of this work.  
Here, a post-processing approach is adopted for which, given an input model of these parameters, the line transfer coefficients can be computed and the (cosmological) 21-cm line radiative transfer calculations can be conducted to predict the observed spectra at individual lines-of-sight. 
For calculations in the cosmological context, the upper limit of $n_{\rm HI}$ is constrained by the baryonic number density and can be further combined with a spin temperature model $T_{s}(z)$ to calculate $n_{\rm{l}}$ or $n_{\rm{u}}$.

\subsection{Code verification}
\label{subsec:codeveri}

A number of numerical tests are conducted  
  to verify 
  the implementation of the algorithm 
  and the execution of the code 
  for C21LRT calculations. 
Appendix~\ref{app:C21LRTCodeVeri} shows the set-up and the numerical results of two example tests that verify the ability of the code 
  to correctly account for the cosmological expansion effects on the transport of continuum radiation. 
Below we summarise what these calculations reveal, in the context of the determination and interpretation of tomographic 21-cm spectra that would be observed, in addition to fulfilling their designs for code verification purposes.

Appendix~\ref{subsec:codeveri-one-CMB} show the cosmological transport of the CMB radiation, as a continuum background, 
in a self-consistent manner as to the transfer of the 21-cm line. 
Machine floating point residuals are obtained, indicating the robustness of the C21LRT numerical scheme for scientific investigations building towards high-precision cosmological studies of reionisation, 
which would require calculations for cosmological radiative transfer of the 21-cm line in the presence of continuum from the Cosmic Dawn and the EoR, up to the present. Such a calculation will be presented in Section~\ref{sec:C21cmStudies}.

Appendix~\ref{subsec:codeveri-two-lineprofile}  
   shows the frequency shift, width compression (in the frequency space) 
   and intensity suppression (due to Lorentz invariance) 
   in cosmological radiative transfer. 
The compression of the line width  
   is significant in the cosmological evolutionary context. 
For instance, the ${\rm FWHM}_{\nu}$ of a line  
  created at $z=10$ 
  would be reduced to a ${\rm FWHM}_{\nu}/(1+z)={\rm FWHM}_{\nu}/11$
  when it is observed at $z=0$.
This width reduction 
  is caused by the expansion of the Universe, 
  and the line width is scaled by $\left[(1+z)/(1+z_{\rm emi})\right]$ (see Appendix~\ref{subsec:codeveri-two-lineprofile} for details), 
  which is the same as the scaling factor of the radiation frequency. 
This scaling will be cancelled out 
  if $\Delta \nu/ \nu$ 
  (where $\Delta \nu$ is the line width) 
  is used in an observational analysis. 
Interpretation of data using a theoretical model 
  is an inverse process. 
Therefore, 
  caution must be taken 
  when using the line width 
  in interpreting spectroscopic results 
  associated with distant sources, 
  and, in particular, 
  the subtleties on cosmological expansion effects 
    and on local physical processes 
    that lead to the change in the line profile    
    must be properly accounted for.

\section{21-cm Tomography with C21LRT}
\label{sec:C21cmStudies}

With the validated ray-tracing code, 
here, the C21LRT equation is solved in representative cosmological settings from $z_{\max}=35.37$ to the present epoch along a single ray, demonstrating the convolution of cosmological effects and radiative transfer effects when they are properly accounted for. 

The diffuse gas suffusing the Universe 
  (hereafter, referred to as the intergalactic medium (IGM) 
  after the appearances of the first luminous structures) 
  consist of two phases:   
(i) ionised (hereafter, referred to as \HII, 
  without losing generality) gas 
 in bubbles embedded with luminous sources that supply the ionising photons, 
(ii) largely neutral IGM in regions 
 outside the \HII bubbles. 
Gas inside the \HII bubbles 
 with a strong radiation field 
 is practically fully ionised,  
 as recombination cannot keep up with ionisation. 
Gas in regions 
  far outside the bubbles 
  would remain neutral.  
Ionised gas and neutral gas 
  can co-exist in a transition region 
  between the \HII bubble 
  and the ambient neutral medium. 
The ionisation state of the gas 
  in these three regions 
  can be described 
  in terms of a parameter,  
  the ionisation fraction $x_{\rm i}$, 
  with $x_{\rm i} = 0$ for neutral gas and   
  and $x_{\rm i} = 1$ for fully ionised gas. 

\begin{figure} 
\centering
\includegraphics[width=0.95\columnwidth]{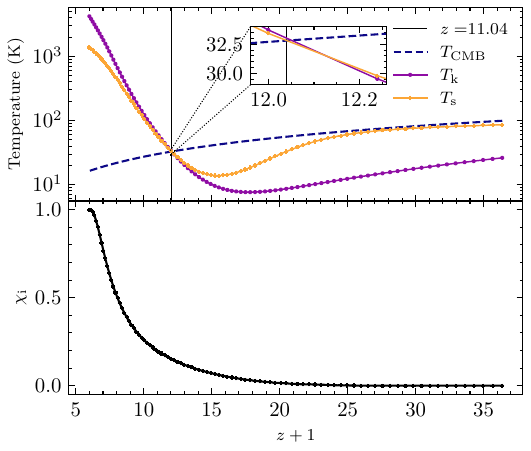}
\caption[Input from 21CMFAST]{
Input properties of \HI gas into C21LRT code based on simulation result of 21CMFAST~\citep{Munoz2022MNRAS}. The globally averaged  kinematic gas temperature $T_{\rm k}$, spin temperature of 21-cm line $T_{\rm s}$ and the CMB temperature $T_{\rm CMB}$ are plotted in the upper panel in purple, yellow and blue (dashed line), respectively. The globally averaged ionisation fraction of \HI $x_{\rm i}$ is plotted in the bottom panel in black. The dots in the lines for $T_{\rm k}$, $T_{\rm s}$ and $x_{\rm i}$ are obtained from 21CMFAST simulation snapshots~\citep{Munoz2022MNRAS}. To match the redshift resolution of our calculation in C21LRT code, we linearly interpolate these quantities between the maximum ($z_{\max}=35.37$) and minimum redshifts ($z_{\min}=5.0$). At $z_{\rm transition}=11.04$, the interpolated $T_{\rm s}$ equals to $T_{\rm CMB}$. The zoomed-in patch shows the region near $z_{\rm transition}$.
}\label{fig:input_21CMFAST}  
\end{figure}  

In this demonstrative study, 
  the detailed structures and the cosmological evolution 
  of these three regions are not considered. 
Instead, the cosmological evolution of the ionisation state 
  of the line-of-sight medium 
  is parametrised by a volume-averaged value 
  for the ionisation fraction, \ie $x_{\rm i}({z})$. 
In other words, the transfer of 21-cm line photons  
  in the \HII bubble, the neutral medium and the transition regions 
  are not explicitly distinguished. 

We adopt the globally averaged $T_{\rm s}(z)$ and $x_{\rm i }(z)$ from the `EOS 2021 all galaxies simulation' result (which used the semi-analytical code 21CMFAST)~\citep{Munoz2022MNRAS} as the inputs of C21LRT calculations, as shown in \Fig\ref{fig:input_21CMFAST}. We also plot $T_{\rm k}$ and $T_{\rm CMB}$ in the upper panel for comparison. To match the redshift resolution of our calculation, we linearly interpolated $T_{\rm s}(z)$ and $x_{\rm i }(z)$ between the maximum ($z_{\max}=35.37$) and minimum redshifts ($z_{\min}=5.0$)~\footnote{The redshift limits are chosen because the simulated results are only available between $z_{\max}$ and $z_{\min}$.}. At $z_{\rm transition}=11.04$, the interpolated $T_{\rm s}$ equals $T_{\rm CMB}$. For $z<5.0$, we consider the remaining \HI negligible for the global 21-cm signal. We let $x_{\rm i}=0$ and do not consider further modifications of the radiations due to \HI at $z<z_{\min}$~\footnote{After the completion of the cosmological reionisation, only the self-shielded surviving dense clumps of \HI would contribute to the cosmological 21-cm line signals. 
The observational imprints 
caused by convolution of these dense \HI structures 
in the post-reionisation era and the ionised bubbles developed in the EoR have not been thoroughly investigated.   
Although these complex issues will not be addressed in the demonstrative calculations here, 
they can been studied explicitly and their observational consequences can be quantified 
  using the C21LRT formulation. }.

In this study, we consider only CMB as the continuum radiation background. Other continuum radiative processes, 
  such as thermal and non-thermal free-free processes
  and synchrotron radiation, 
  are ignored. 
  Then based on the input $T_{\rm s}(z)$, we expect to see two regimes: one for absorption and one for emission. Between $z_{\max}$ and $z_{\rm transition}$, the competition between radiative coupling with the CMB and the collisional coupling leads to $T_{\rm CMB} \gtrsim  T_{\rm s} > T_{\rm k}$, and the 21-cm signal is in an absorption regime. 
Afterwards, the first luminous sources released UV radiation and X-rays which ionised and heated the \HI gas. This ushered the Universe 
into the cosmological reionisation epoch.
More importantly, the UV radiation would 
give rise to \Lya pumping process, 
and the hyperfine states of the \HI gas 
were no longer determined 
solely by the thermal and collision coupling processes. 
The \Lya radiation boosted the relative populations of \HI atoms 
in the upper (triplet) hyperfine state via the Wouthuysen-Field mechanism, 
while the heating by UV radiation and X-rays of the \HI gas 
resulted in $T_{\rm k} > T_{\rm r}$. Hence the 21-cm signal is in an emission regime between $z_{\rm transition}$ and $z_{\min}$.

With $x_{\rm i}(z)$ specified, 
  the amount of \HI gas 
  and the relative population of \HI atoms in the two hyperfine states 
  can be computed given a cosmological model, 
  with the procedures described as follows. 
The density of hydrogen is approximately 75\% 
  of the baryon density by mass, 
  \ie $\rho_{\rm H}=3\rho_{\rm b}/4$, 
  and the remaining 25\% is mainly contributed by helium, 
  especially before the mass production of metals in stars. 
Assuming a two-species (H-He) description of the baryonic content,  
  the cosmological number density of hydrogen is then    
\begin{align}
\overline{n}_{\rm H}(z)  
   & = 
  \left(\Omega_{\rm b}(z) / m_{\rm p}\right) 
     \rho_{\rm crit}\left(1-Y_{\rm He}\right) \nonumber \\ 
   & \approx 
     \left(\Omega_{{\rm b}, 0} / m_{\rm p}\right)(1+z)^3 
     \rho_{\rm crit}\left(1-Y_{\rm He}\right) \nonumber \\ 
   & = 
    \overline{n}_{{\rm H}, 0} (1+z)^3  \ .   
\label{eq:nHIevo}
\end{align} 
In the calculations, 
  the mass fraction of helium is set to be 
  $Y_{\rm He} = 1/4$. 
The present baryonic density  
  $\rho_{{\rm b}, 0} = 4.18977 \times  10^{-31}~{\rm g~cm}^{-3}$, 
  deduced from 
   $\Omega_{{\rm b}, 0} =\rho_{{\rm b}, 0}/\rho_{\rm crit}$, 
    where $\Omega_{{\rm b}, 0}h^2 = 0.02230$ 
    \citep{Planck2015_CosmoPara},  
     $\rho_{\rm crit} = 3H_0/(8\pi G) 
      = 1.87882 \times 10^{-29} h^2~{\rm g~cm}^{-3}$ 
      (with $h = 0.6774$), 
    and hence, the present hydrogen number density  
   $\overline{n}_{\rm H, 0} = 1.87745 \times 10^{-7}~{\rm cm}^{-3}$. 
It follows that 
\begin{align} 
  {\overline{n}_{\rm HI}}(z) 
    & = \overline{n}_{\rm H}(z)\, x_{\rm HI}(z) 
    = \overline{n}_{\rm H}(z)\,(1-{x_{\rm i}}(z))   \  .  
\end{align}

For each ray traced in our calculations, the propagation of the radiation along the ray is stamped by the redshift $z$. 
The number density of the \HI atoms 
in the lower (singlet) hyperfine state along one ray is given by    
\begin{align}
   n_{\rm l}(z) 
  & =   \frac{n_{\rm HI}(z)}
      {1+3\,{\rm exp}(-T_{\star}/{T_{\rm s}(z)})}\approx 
      \frac{n_{\rm HI}(z)}
     {4-3 (T_{\star}/{T_{\rm s}(z)})}   
       \nonumber \\
   & =  
       \frac{\overline{n}_{\rm H}(z) x_{\rm HI}(z)
      \left(1+\delta_{\rm b}(z)\right)}
      {4-{3\,T_{\star}}/{T_{\rm s}(z)}},
\end{align} 
for $T_{\rm s} \gg T_{\star}$, 
where the baryonic matter overdensity is 
$\delta_{\rm b} \equiv (\rho_{\rm b} /\overline{\rho}_{\rm b} -1)$, with  $\overline{\rho}_{\rm b}$ being the mean density. 
The corresponding number density of the \HI atoms 
  in the upper (triplet) hyperfine state is given by  
   ${n_{\rm u}}(z) = n_{\rm HI}(z) - n_{\rm l}(z)$. 
Therefore, 
\begin{align}
 n_{\rm u}(z) & \approx
  \frac{3\,\overline{n}_{\rm H}(z) x_{\rm HI}(z)
   \left(1+\delta_{\rm b}(z)\right) \,(1-T_{\star}/T_{\rm s}(z))}
   {4-3\,T_{\star}/T_{\rm s}(z)}. 
\end{align} 
The above parametrisation 
  of the relative population of \HI atoms in the two hyperfine states, 
  $(n_{\rm u}/n_{\rm l})$,  
  using the spin temperature $T_{\rm s}$   
  has imposed 
  a constant $3:1\: (=g_{\rm u}:g_{\rm l})$ 
  ratio for the relative populations\footnote{
Although beyond the scope of this study, we note that 
this parametrisation 
  will be invalid  
  if there is a strong UV radiation field. 
The \Lya pumping will allow 
  a higher number of \HI atoms 
  in the upper hyperfine state  
  through the Wouthuysen-Field mechanism. 
In this situation, 
  an additional local radiative transfer calculation 
  will be required 
  so to determine $n_{\rm u}$ and $n_{\rm l}$.}

\begin{figure}
\centering
\includegraphics[width=1.0\columnwidth]{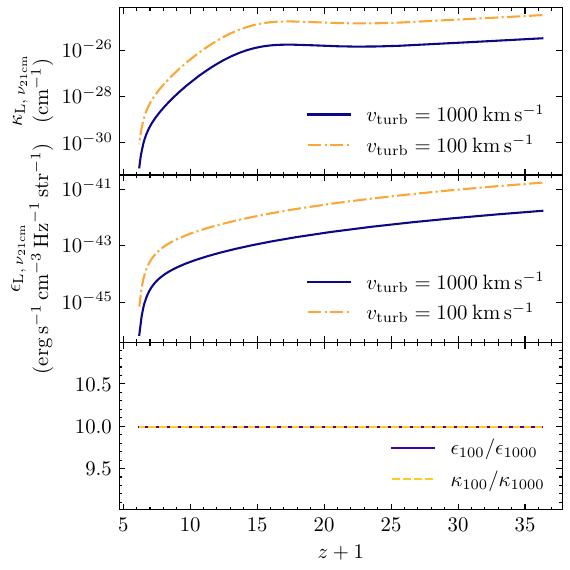}
\caption[coeff]{The absorption and emission coefficients ($\kappa_{\rm L, \nu21cm}$ and $\epsilon_{\rm L, \nu21cm}$) 
of 21-cm line 
at the frequency $\nu = \nu_{21{\rm cm}}$ 
evaluated in the comoving frame at $z$. Coefficients calculated with  
 $v_{\rm turb}=1000\:{\rm km}\,{\rm s}^{-1}$ 
and $100\:{\rm km}\,{\rm s}^{-1}$ are plotted
 in the panel on top and in the middle, respectively. These coefficients
 are calculated within the C21LRT code and saved along with other results. 
The ratio of absorption and emission coefficients when adopting $v_{\rm turb}=100\:{\rm km}\,{\rm s}^{-1}$ and $1000\:{\rm km}\,{\rm s}^{-1}$ 
are plotted in the bottom panel, with solid purple line
 and dashed yellow line, respectively. 
Both their values are
exactly 10 determined by the (normalised) line profile as discussed in Sec \ref{subsubsec:LineProfile}. 
}
\label{fig:emi_abs_coeff_C21LRT}  
\end{figure} 
With the properties of \HI specified,  we still need to specify the line profile of 21-cm line to determine the 21-cm radiative processes.
Line broadening caused by thermal motion 
  is insignificant 
  in this demonstrative study\footnote{To match the broadening caused by $v_{\rm turb}=10\,{\rm km s^{-1}}$, we need $T_{\rm k}=6060.67$~K, which is already higher than expected $T_{\rm k}$ of \HI as shown in \Fig\ref{fig:input_21CMFAST}.}.  
The line broadening 
  is, therefore, caused only by turbulent motion 
  characterised by a root-mean-square velocity 
  $v_{\rm turb}$.  
Furthermore, $v_{\rm turb}$ is assumed to be uniform along the line-of-sight and constant through out one set of calculation. We do not consider the other types of relative velocities of the \HI gas along line-of-sight and leave redshfit space distortion effects to future studies.

\section{Results and Discussion}\label{sec:results}

\subsection{Line transfer coefficients}\label{subsec:linetransfercoef}
At a given $z$, 
the variation of the transfer coefficients across the line frequency $\nu$ follows 
the line profile function $\phi_\nu$. \fig{\ref{fig:emi_abs_coeff_C21LRT}} 
 shows the transfer coefficients of the \HI hyperfine line 
 (without a continuum) at the line centre, 
 evaluated in the comoving $z$ frames 
 for the case with $v_{\rm turb} = 1000\;{\rm km\,s}^{-1}$ 
 and the comparison with the cases 
 with $v_{\rm turb} = 100\;{\rm km\,s}^{-1}$. The emission and absorption coefficients at the line centre $\kappa_{{\rm L},\nu{\rm 21cm}}$ and $\epsilon_{{\rm L},\nu{\rm 21cm}}$ are proportional to $1/v_{\rm turb}$ as discussed in Sec \ref{subsubsec:LineProfile}, shown in the bottom panel. As the corresponding FWHM$_{\nu}$ is relatively small compared to the entire frequency range in the 21-cm spectra at $z=0$, even when we adopt $v_{\rm turb} = 1000\;{\rm km\,s}^{-1}$, and the 21-cm line profile is well resolved in frequency space, we do not expect the difference to show up in the 21-cm global spectra calculated with smooth 
 \HI properties in this paper.
The differences due to different $v_{\rm turb}$ is significant when the properties of \HI changes on scales comparable to or smaller than FWHM$_{\nu}$, as analysed in~\citet{Wu2023MNRAS}.

\subsection{Tomographic 21-cm Spectra}
\label{subsec:linespectra}
The radiative transfer of the 21-cm hyperfine line 
  from $z_{\max}$ 
  to $z=0$ (the present epoch) 
 is solved using a ray-tracing approach with the input specified in Sec~\ref{sec:C21cmStudies} (see \Fig\ref{fig:input_21CMFAST}), while adopting various turbulent velocities. 
  
\begin{figure} 
\centering
\includegraphics[width=1.0\columnwidth]{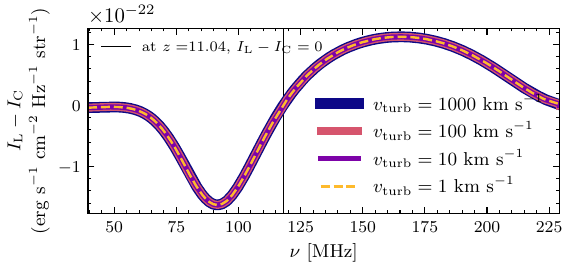}
\caption[21-cm spectra at z=0 with various turbulent velocity]{ Spectra of 21-cm line at $z=0$ using the inputs from \Fig\ref{fig:input_21CMFAST}, with various assumed turbulent velocity. The spectra calculated with $v_{\rm turb}= 1000, 100, 10, 1~{\rm km~s^{-1}}$ are plotted in blue (thickest, in the bottom layer), crimson, purple and yellow (thinnest, in the top layer, with dashed line). These lines are perfectly overlapped all over the plotted frequency range when inspected by eye, as expected (see detailed explanation in main text). The relative difference between these four scenarios are small (see Appendix~\ref{app:C21LRTAlgo} for details.). The intensity of 21-cm line (after subtracting CMB continuum) changes from negative to positive at $\nu=118.97$ MHz, which corresponds to $z=11.04$ where $T_{\rm s}=T_{\rm CMB}$ in \Fig\ref{fig:input_21CMFAST} . 
}
\label{fig:21_spectra_z0_various_v_turb}  
\end{figure} 
\begin{figure}
\centering
\includegraphics[width=1.0\columnwidth]{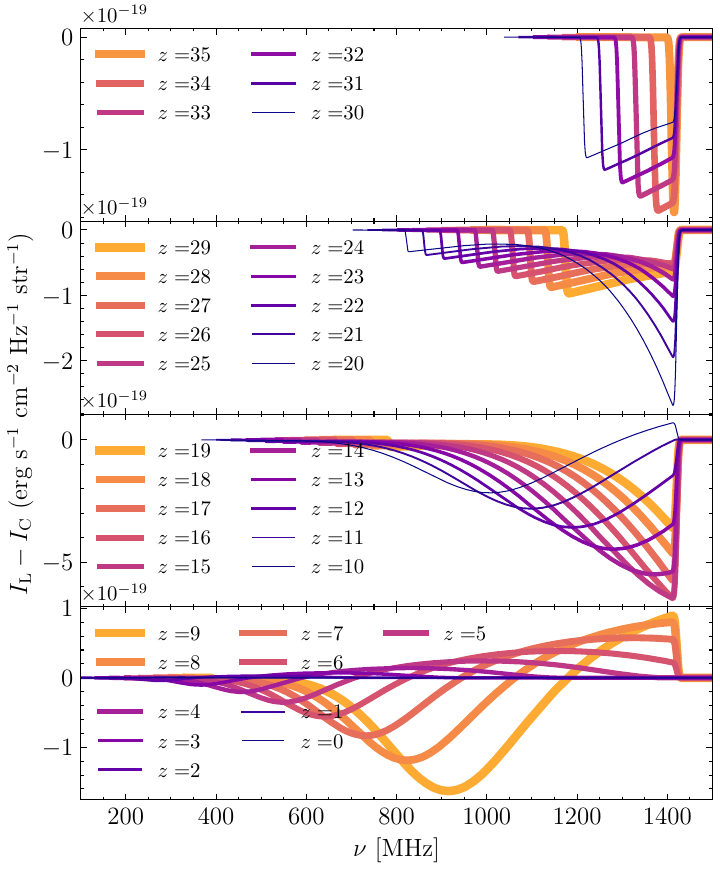}
\caption[C21LRT demonstrative study results: 
21-cm line spectra in the comoving frames at redshifts $35.0 \geq z \geq 0$ for the case with $b_{\rm D} = v_{\rm turb} =1000\:{\rm km}\,{\rm s}^{-1}$]
{Spectra of the 21-cm line in the rest frames of $z=35, 34, ..., 0$ when the C21LRT calculated has been carried out to the corresponding redshifts. In each panel, higher-redshift results are plotted with thicker lines. The spectrum corresponding to the highest and lowest redshift in a panel is coded in orange and purple, respectively. For all spectra, the y-value (specific intensity) collapses to 0 quickly near $\nu_{\rm 21cm}$. This is because the radiative transfer calculation has only been carried out to a certain redshift. In the rest frame of this redshift, the amount of emission blueward of $\nu_{\rm 21cm}$ is negligible. Also, the strength of the signal is reduced from higher to lower redshift as the result of cosmological expansion. This effect is most noticeable in the bottom panel, where the spectra corresponding to the lowest redshift values have almost collapsed to 0 (see \Fig\ref{fig:21_spectra_z0_various_v_turb} $I_{\rm L}-I_{\rm C}\sim 10^{-22}{\rm erg\,s^{-1}\,cm^{-3}\,Hz^{-1}\,str^{-1}}$  at $z=0$). 
     } 
\label{fig:21_spectra_development} 
\end{figure}

\begin{figure}
\centering
\includegraphics[width=1.0\columnwidth]{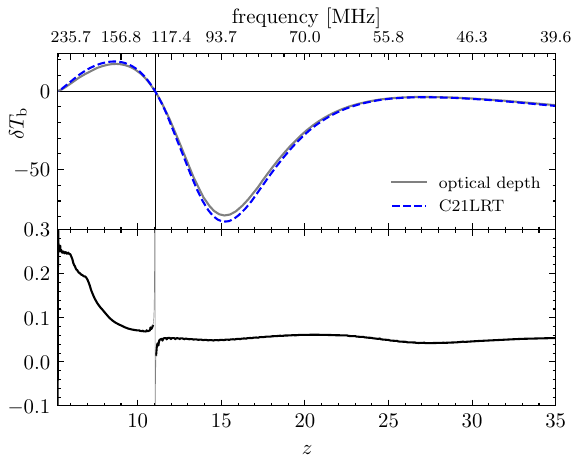}
\caption[caseII]
{Comparing the results from C21LRT calculations and optical method on 21-cm differential brightness temperature $\delta T_{\rm b}$ - redshift $z$ diagram. In the upper panel, $\delta T_{\rm b}$ calculated with C21LRT and optical depth parametrisation (\eqn{\ref{eq:delta_Tb}}) are plotted in blue dashed line and grey solid line, respectively. The relative difference $\delta T_{\rm b~C21LRT}/\delta T_{\rm b~optical~depth}-1$ is plotted in the lower panel. The region around $z_{\rm transition}=11.04$ is made less apparent intentionally because this is where $\delta T_{\rm b}$ is very close to zero near $z_{\rm transition}$ and the relative difference increases artificially. In the upper panel, $z_{\rm transition}$ and $\delta T_{\rm b}=0$ are marked with think black lines.}
\label{fig:compare} 
\end{figure}


Spectra of 21-cm line at $z=0$ calculated with $v_{\rm turb}= 1000,\ 100,\ 10,\ 1~{\rm km\,s^{-1}}$ are shown in \Fig\ref{fig:21_spectra_z0_various_v_turb} 
with thicker lines corresponding to larger $v_{\rm turb}$ ( $v_{\rm turb}=1000~{\rm km\,s^{-1}}$ in blue with the thickest line in the bottom layer, $v_{\rm turb}=100,\ 10~{\rm km\,s^{-1}}$ in crimson and purple, $v_{\rm turb}=1~{\rm km\,s^{-1}}$ in yellow with thinnest dashed line in the top layer). These spectra are perfectly overlapped over the entire plotted frequency range when inspected by eye, as expected. 
Quantitatively, the relative difference between these four scenarios are small, on order of magnitudes of $10^{-5} - 10^{-4}$ (see Appendix~\ref{app:C21LRTAlgo} for details). The intensity of 21-cm line (after subtracting CMB continuum) changes from negative to positive at $\nu=118.973$ MHz, which corresponds to $z=11.04$ where $T_{\rm s}=T_{\rm CMB}$ in \Fig\ref{fig:input_21CMFAST}.

We track the development of the prominent spectral features and how they change in amplitude and shift in redshift space as the radiation propagated down to $z=0$. We show spectra in the local rest frames at $z=35, 34, ..., 0$ in \Fig~\ref{fig:21_spectra_development}. In each panel, higher redshift results are plotted with thicker lines. For example, in the top panel, the spectra corresponding to $z=35$ and $z=30$ are plotted with the thickest orange and thinnest purple lines, respectively. For all spectra, the y-value (intensity) collapses to 0 quickly near $\nu_{\rm 21cm}$. 
This is because the calculation has only been carried out to a certain redshift. In the rest frame of this redshift, the amount of emission blueward of $\nu_{\rm 21cm}$ is negligible. The strength of the 21-cm signal is reduced from higher to lower redshift as the consequence of cosmological expansion. This effect is most noticeable in the bottom panel. The maximum negative intensity is $I_{\rm L}-I_{\rm C}\sim 10^{-19}{\rm erg\,s^{-1}\,cm^{-3}\,Hz^{-1}\,str^{-1}}$ for $z=9$, which can be compared to the spectra in \Fig~\ref{fig:21_spectra_z0_various_v_turb} where $I_{\rm L}-I_{\rm C}\sim 10^{-22}{\rm erg\,s^{-1}\,cm^{-3}\,Hz^{-1}\,str^{-1}}$  at $z=0$.

\subsection{Compare C21LRT with the optical-depth parametrisation}
\label{subsec:optical_depth_Discussion} 

Current calculations of 
21-cm tomographic spectra 
often use the following expression: 
\begin{align} 
\delta T_{\mathrm{b}} 
& \approx 
\left( 
\frac{T_{\mathrm{s}}-T_{\mathrm{r}}}{1+z}
\right)\;\! \tau_{\nu} \nonumber \\
& \approx  27\  x_{\mathrm{HI}}\left(1+\delta_{\mathrm{b}}\right)\left(\frac{\Omega_{\mathrm{b}} h^{2}}{0.023}\right)
\left(\frac{0.15}{\Omega_{\mathrm{m}} h^{2}}\right)^{1/2}  \left(\frac{1+z}{10}\right)^{1 / 2}  \nonumber \\
& \hspace*{1cm}
\times\left(\frac{T_{\mathrm{s}}-T_{\mathrm{r}}}{T_{\mathrm{s}}}\right)\left[\frac{\partial_{\mathrm{r}} v_{\mathrm{r}}}{(1+z)\;\! H(z)}\right] \mathrm{mK} 
\label{eq:delta_Tb}
\end{align}  
\citep[see][]{Furlanetto2006_review, Pritchard2012Review, Abdurashidova2022ApJ_HERA}.  
Here, $T_{\mathrm{r}}$ is the brightness temperature 
  of the background radiation at the relevant frequency, 
$x_{\mathrm{HI}}$ is the neutral fraction of hydrogen, 
$\delta_{\mathrm{b}}$ is the fractional over-density in baryons, 
and ${\partial_{\mathrm{r}} v_{\mathrm{r}}}$ 
  specifies the gradient of proper velocity along the line-of-sight.   
The expression was derived in~\cite{Furlanetto2006_review}, where various 
approximations 
were taken to obtain such an analytical expression. Most importantly, they assumed that 
(i) \HI gas is moving uniformly with the Hubble flow then used this assumption to skip distance integration (length of the \HI gas) and frequency integration (frequency range defined by 21-cm line profile at a given distance)~\footnote{An alternative derivation and detailed interpretations of Eqn.~(\ref{eq:delta_Tb}) can also be found in~\cite{Pritchard2012Review}.}. 
This essentially wipes out the contribution of variations in $n_{\rm HI}$ and $T_{\rm s}$ on small scales to the variations in the 21-cm signal. 

The other minor approximations adopted while deriving \eqn{\ref{eq:delta_Tb}} were 
(ii) the values of the physical constants or quantities are truncated at only a few digits (e.g. the value of Einstein coefficient), 
(iii) stimulated emission is approximated in first order (of spin temperature $T_{\rm s}$), 
(iv) local thermodynamics equilibrium (LTE) is always assumed (see \eqn{\ref{eq:approx_LTE}} and discussion in Sec \ref{subsec:lineTransCoef}), and 
(v) radiative transfer effects were assumed to be insignificant.

We calculated the 21-cm differential brightness temperature $\delta T_{\rm b}$ with Eqn.~(\ref{eq:delta_Tb}) and used the same inputs ($x_{\rm i}$ and $T_{\rm s}$) as well as cosmological parameters. The result is shown in \Fig~\ref{fig:compare} as indicated with the solid grey line in the upper panel. For comparison, the result obtained with C21LRT (the same in \Fig\ref{fig:21_spectra_z0_various_v_turb}) is shown in the same figure with the blue dashed line. Their relative difference $\delta T_{\rm b~C21LRT}/\delta T_{\rm b~optical~depth}-1$ is plotted in the lower panel. After verifying the differences caused by the minor approximations (ii)-(v) one by one, we found that they all leads to differences less than 1\%. We conclude that the $\gtrsim 5\%$ level difference in the lower panel is caused by the major approximation (i), which is to invoke Hubble flow and skip distance and frequency integration. The relative difference seems to be evolving with redshift and increases quickly after $z < 11$. We therefore tested the discrepancy between Eqn.~(\ref{eq:delta_Tb}) and C21LRT with arbitrary $T_{\rm s}$. We found that the relative difference always increases towards lower redshift regardless of the input $T_{\rm s}$. We tentatively conclude that this is because the major approximation (which used the Hubble flow to skip integration) deviates more from actual distance and frequency integration calculations at lower redshift. We note that the $\gtrsim 5\%$ level relative difference already warrant more investigation and implies that we may derive a more accurate analytical expression with approximate distance and frequency integration. 

In the context of 21-cm tomography, 
this approximation implies that 
the observed $\delta T_{\rm b}$ at $z = 0$ at a specific frequency $\nu_{0}$ is unambiguously connected to frequencies $\nu\vert_z = \nu_{0}\;\!(1+z)$  
along the-line-sight in the redshift space (as in the ray-tracing shown in \Fig~\ref{fig:C21LRTRectGridRayTrace}), with contribution weighted by the opacity appropriate for line-of-sight radiation field and media. 
The one-to-one connection between $\nu_0$ and $\nu\vert_z$ along the line-of-sight would break down,  
when there is a frequency spread in the 21-cm line, induced by local thermal and dynamical properties of the line-of-sight media. 
This frequency spread will not pose a very serious problem  
when the line is optically thin 
and when there is no convolution 
between the broadened 21-cm line 
and its neighbouring continuum 
in the radiative transfer,  
i.e. the expression in Eqn.~(\ref{eq:delta_Tb}) is applicable. 
However, in real situations, we usually have a 21-cm line broadened by the local dynamics, such as bulk motion and turbulence. The time (redshift) ordering and the frequency ordering become entangled. 
Because of the convolution in the radiative transfer processes, the line signal in the 21-cm spectrum at a particular frequency $\nu(z)$ when observed at $z=0$ will not properly reflect the physical conditions of the Universe at redshift $z$.

Note also that the $\gtrsim 5\%$ level relative difference pertains to a simple scenario of a smooth (globally averaged) model of reionisation and that the radio continuum background is solely attributed to the CMB. 
A more substantial difference is anticipated when considering the heterogeneous aspects of reionisation. 
For example, $n_{\rm HI}$ and/or $T_{\rm s}$ may change drastically at the boundaries of ionised zones created by luminous sources. Predicting $\delta T_{\rm b}$ for these regions based on Eqn.~(\ref{eq:delta_Tb}) would create corresponding sharp changes in the 21-cm signal, while in reality 21-cm line is likely to be broadened and a sudden change in the \HI gas properties does not necessarily lead to sharp changes in 21-cm spectra (see~\citet{Wu2023MNRAS}). 
Furthermore, we have not evaluated additional factors contributing to the continuum background, such as emissions from bright radio sources. 
The effects of absorption and emission of continuum along the line-of-sight could further complicate the interplay between the line-continuum interactions, which is further discussed in Appendix~\ref{app:LineContinuumDiscussion}. 
Evaluating the impacts of these factors on the spectra need to be done in relation to their specific contexts. Such an assessment, which extends beyond the scope of this paper, is planned for our future research. 
We intend to use the C21LRT formulation, which is constructed without making specific assumptions about the radiation background, ionisation sources, line-of-sight media, or the history of reionisation, to 
(i) study the impact of these physical factors on the spectra, 
(ii) predict not only the spectra but also power spectra of 21-cm signals for more realistic, patchy reionisation scenarios, 
and, as an inherent bonus of this approach, 
(iii) perform quantitative assessment of the accuracy of the common optical depth parametrisation method. 

\subsection{Observational perspective}

While only single-ray C21LRT calculations were presented along with a focused discussion on the cosmological radiative transfer effects and their implications on 
the global tomographic 21-cm signals that would be seen by the observer at $z=0$, which are targeted by experiments such as EDGES \citep[][]{Bowman2018Nature_EDGES}, 
LEDA \citep[][]{Price2018MNRAS_LEDA}, 
SARAS \citep[][]{Singh2018ExA_SARA2}, 
SCI-HI \citep[][]{Voytek2014ApJ_SCIHI} and 
Prizm \citep[][]{Philp2019JAI_PRIZM}, 
employing multiple-ray (i.e. pencil beam or all sky) C21LRT calculations 
will yield 
cosmological 21-cm images or fluctuations maps. 
The C21LRT formulation is able to take the advantage of the vast resources in the sophisticated reionisation simulations through interfacing the C21LRT calculations with simulations via a post-processing approach, where simulation outputs that provide the redshift and spatial evolution of various 21-cm line related quantities (e.g. number density of \HI, spin temperature, thermal history of the IGM) are sourced to compute the transfer coefficients and the C21LRT equations, and by adopting an appropriate ray-tracing algorithm to solve the transfer equations, to obtain high-fidelity 21-cm spectra templates. 
Statistical properties of the observables in these maps can then be determined \citep[e.g.][]{Watkinson2014MNRAS, Watkinson2017MNRAS, Suman2018MNRAS, Chen2019ApJ, Watkinson2019MNRAS, Greig2022MNRAS, Choudhury2022MNRAS} for drawing physical interpretations of observations and 
understanding of the cosmological reionisation, such as by 
LOFAR\citep[][]{vanHaarlem2013AA_LOFAR}, 
MWA \citep[][]{Tingay2013PASA_MWA}, 
HERA \citep[][]{DeBoer2017PASP_HERA} and 
SKA \citep[][]{Mellema2015aaska, Koopmans2015aska}. 

\section{Conclusion}\label{sec:conclusion}

The observed spectral properties of 21-cm line radiation associated with cosmological reionisation are modified by both cosmological and astrophysical processes that 
span broad ranges of length scales 
and time scales over cosmic time. 
A covariant formalism for cosmological radiative transfer 
 of the 21-cm line was devised 
  from the first principles, 
  based on conservation of phase-space volume of the radiation 
  and the conservation of photon number. 
In the formulation, 
  the local radiation processes are included explicitly   
  in terms of the absorption and emission coefficients 
  of the line and the continuum, 
  and the local line broadening in terms of 
  a multiplicative line profile function 
  in the line transfer coefficients.  
  
From this covariant formulation, 
  a C21LRT equation 
  was derived assuming FLRW space-time. 
A ray-tracing algorithm is adopted for the numerical code 
  to solve the C21LRT equation. 
The code takes full account of 
  global effects associated with cosmological evolution,   
  local effects on line broadening and frequency shifting 
  and the convolution caused by line and continuum radiative transfer.  
The code is verified by a battery of numerical tests:  
  including the cosmological transfer of CMB continuum,  
  the transfer of a generic line in an expanding universe, 
  and the transfer of a 21-cm line in the presence of  
  absorption in structured media 
  with differential rotational motions. 

We present a set of demonstrative calculations 
and generate tomographic 21-cm spectra 
using the reionisation history based on~\citep{Munoz2022MNRAS}.  
We showed how the 21-cm spectral signal developed and propagated throughout EoR in C21LRT. We adopted Gaussian 21-cm line profiles determined by turbulent velocities and found that these Gaussian line profiles cause negligible difference to the 21-cm signal spectra at $z=0$. This is mainly because we used the globally averaged \HI gas properties, which are smoothly varying with redshift. 

Compared the spectrum at $z=0$ 
with that obtained by the optical depth parametrisation (Eqn.\ref{eq:delta_Tb}, \citet{Pritchard2012Review}), 
we have found the discrepancy between the $\delta T_{\rm b}$ 
at the level of $5\%$ for redshift $z \sim 12$ or higher 
and of $>10\%$ for redshift smaller than $z\sim 8$.

The C21LRT formulation's independence from specific assumptions regarding radiation background, ionisation sources, line-of-sight media, reionisation history, and the 21-cm line profile, enhances its utility in accurately predicting 21-cm signals during reionisation. 
We have demonstrated that our C21LRT formulation 
  can explicitly and self-consistently treat 
  (i) time (redshift) and frequency ordering, 
  (ii) line-continuum interactions 
  in the presence of emission and absorption,  
  and (iii) convolution 
  of line and continuum radiative transfer 
  with cosmological expansion.   
Within the confines of this paper, we have focused on applying the C21LRT calculation to a globally smooth reionisation model and considering a radio continuum background solely consisting of the CMB. 
Future research building on the C21LRT formulation will delve deeper into the heterogeneous nature of reionisation, exploring how it impacts the observed spectra in more intricate scenarios. 
For instance, in \citet{Wu2023MNRAS}, we have investigated the redshifted 21-cm imprints across ionised cavities in an expanding universe, accounting for factors such as differing ionised bubbles sizes at different redshifts, as well as the distortion of the apparent shape of the ionisation front due to the finite speed of light. 
Additionally, upcoming investigations will look into the impacts on the 21-cm spectra in scenarios where a continuum component, in additional to the CMB, is present, such as emissions from bright radio quasars and galaxies. 
These studies will also investigate the 21-cm signals, 
both in spectra and power spectra, 
considering the diverse, complex dynamics that occur during reionisation across different cosmic environments and redshifts. 
Quantitative comparison will be drawn against the optical depth parametrisation method that often resorts to restrictive assumptions and approximations for analytical tractability but overlooks line and continuum radiative transfer effects and small-scale variations in the neutral hydrogen density and the spin temperature, as discussed in Section \ref{subsec:optical_depth_Discussion}. 
These research efforts underscore the importance 
of proper covariant radiative transfer  
   for computing model 21-cm spectra 
for extracting meaningful information 
regarding the structural development of the Universe 
from the (all-sky) 21-cm tomographic data, 
  from observations such as those by MWA, HERA and SKA.

\section*{Acknowledgements}
We thank 
Jonathan Pritchard, 
Jacqueline Hewitt, 
Adrian Liu, 
Jordan Mirocha, 
Emma Chapman, 
Catherine Watkinson, 
Geraint Harker, 
Girish Kulkarni, 
Suman Majumdar, 
for discussions 
  on astrophysics of the Epoch of Reionisation
  and 21-cm tomography,  
  Thomas Kitching on the 
  cosmological and astrophysical 
  aspects of this study,
  John Richard Bond and Juna Kollmeier on treatments of transport processes, 
  and Ziri Younsi on numerical covariant radiative transfer. 
  JYHC acknowledges the enriching discussions and knowledge exchange regarding the theoretical and observational facets of reionisation at the forums of the Pan-Canadian CITA Reionisation Focus Group and the CMB+EoR Summer Workshop at McGill University. 
JYHC was supported by the Natural Sciences and Engineering Research Council 
  of Canada (NSERC) [funding reference \#CITA 490888-16] 
  by a CITA Postdoctoral Fellowship, 
  as well as the University of Toronto Faculty of Arts \& Science Postdoctoral Fellowship with the Dunlap Institute for Astronomy \& Astrophysics. 
  The Dunlap Institute is funded through an endowment established by the David Dunlap family and the University of Toronto. 
QH was supported by UCL through an Overseas Research Scholarship 
  and the UK STFC through a postgraduate research studentship. 
QH, KW and JDM acknowledge the support 
  from the UCL Cosmoparticle Initiative. 
This work was partially supported by a 
  UK STFC Consolidate Grant awarded to UCL-MSSL. 
This work had made use of NASA’s Astrophysics Data System.

\section*{Data Availability}

The theoretical data generated in the course of this study are available from the lead author, upon reasonable request.
 



\bibliographystyle{mnras}
\bibliography{main} 

\begin{thebibliography}{}
\makeatletter
\relax
\def\mn@urlcharsother{\let\do\@makeother \do\$\do\&\do\#\do\^\do\_\do\%\do\~}
\def\mn@doi{\begingroup\mn@urlcharsother \@ifnextchar [ {\mn@doi@} {\mn@doi@[]}}
\def\mn@doi@[#1]#2{\def\@tempa{#1}\ifx\@tempa\@empty \href {http://dx.doi.org/#2} {doi:#2}\else \href {http://dx.doi.org/#2} {#1}\fi \endgroup}
\def\mn@eprint#1#2{\mn@eprint@#1:#2::\@nil}
\def\mn@eprint@arXiv#1{\href {http://arxiv.org/abs/#1} {{\tt arXiv:#1}}}
\def\mn@eprint@dblp#1{\href {http://dblp.uni-trier.de/rec/bibtex/#1.xml} {dblp:#1}}
\def\mn@eprint@#1:#2:#3:#4\@nil{\def\@tempa {#1}\def\@tempb {#2}\def\@tempc {#3}\ifx \@tempc \@empty \let \@tempc \@tempb \let \@tempb \@tempa \fi \ifx \@tempb \@empty \def\@tempb {arXiv}\fi \@ifundefined {mn@eprint@\@tempb}{\@tempb:\@tempc}{\expandafter \expandafter \csname mn@eprint@\@tempb\endcsname \expandafter{\@tempc}}}

\bibitem[\protect\citeauthoryear{{AlOmar}}{{AlOmar}}{2020}]{AlOmar2020VoigtProfile}
{AlOmar} S.~A.,  2020, \mn@doi [Optik] {https://doi.org/10.1016/j.ijleo.2019.163919}, 203, 163919

\bibitem[\protect\citeauthoryear{{Bacon} et~al.,}{{Bacon} et~al.}{2021}]{Bacon2021AA_MUSE}
{Bacon} R.,  et~al., 2021, \mn@doi [\aap] {10.1051/0004-6361/202039887}, \href {https://ui.adsabs.harvard.edu/abs/2021A&A...647A.107B} {647, A107}

\bibitem[\protect\citeauthoryear{{Barkana} \& {Loeb}}{{Barkana} \& {Loeb}}{2001}]{BarkanaLoeb2001PhR}
{Barkana} R.,  {Loeb} A.,  2001, \mn@doi [\physrep] {10.1016/S0370-1573(01)00019-9}, \href {https://ui.adsabs.harvard.edu/abs/2001PhR...349..125B} {349, 125}

\bibitem[\protect\citeauthoryear{{Bennett} et~al.,}{{Bennett} et~al.}{2013}]{Bennett_13_9yrsWMAP}
{Bennett} C.~L.,  et~al., 2013, \mn@doi [ApJS] {10.1088/0067-0049/208/2/20}, \href {http://adsabs.harvard.edu/abs/2013ApJS..208...20B} {208, 20}

\bibitem[\protect\citeauthoryear{{Bezanson} \& {Franx}}{{Bezanson} \& {Franx}}{2012}]{Bezanson2012GalaxiesVeloDisp}
{Bezanson} R. d~{van Dokkum} P.,  {Franx} M.,  2012, \mn@doi [\apj] {10.1088/0004-637X/760/1/62}, \href {https://ui.adsabs.harvard.edu/abs/2012ApJ...760...62B} {760, 62}

\bibitem[\protect\citeauthoryear{{Bond}, {Kofman}  \& {Pogosyan}}{{Bond} et~al.}{1996}]{Bond1996Nature}
{Bond} J.~R.,  {Kofman} L.,   {Pogosyan} D.,  1996, \mn@doi [\nat] {10.1038/380603a0}, \href {https://ui.adsabs.harvard.edu/abs/1996Natur.380..603B} {380, 603}

\bibitem[\protect\citeauthoryear{{Bowman}, {Rogers}, {Monsalve}, {Mozdzen}  \& {Mahesh}}{{Bowman} et~al.}{2018}]{Bowman2018Nature_EDGES}
{Bowman} J.~D.,  {Rogers} A. E.~E.,  {Monsalve} R.~A.,  {Mozdzen} T.~J.,   {Mahesh} N.,  2018, \mn@doi [\nat] {10.1038/nature25792}, \href {https://ui.adsabs.harvard.edu/abs/2018Natur.555...67B} {555, 67}

\bibitem[\protect\citeauthoryear{Boyer \& Lynas-Gray}{Boyer \& Lynas-Gray}{2014}]{Boyer14VoigtEval}
Boyer W.,  Lynas-Gray A.~E.,  2014, \mn@doi [\mnras] {10.1093/mnras/stu1606}, 444, 2555

\bibitem[\protect\citeauthoryear{{Breysse} et~al.,}{{Breysse} et~al.}{2022}]{Breysse2022ApJ_COMAP}
{Breysse} P.~C.,  et~al., 2022, \mn@doi [\apj] {10.3847/1538-4357/ac63c9}, \href {https://ui.adsabs.harvard.edu/abs/2022ApJ...933..188B} {933, 188}

\bibitem[\protect\citeauthoryear{{Carilli}}{{Carilli}}{2011}]{Carilli2011ApJ_CO}
{Carilli} C.~L.,  2011, \mn@doi [\apjl] {10.1088/2041-8205/730/2/L30}, \href {https://ui.adsabs.harvard.edu/abs/2011ApJ...730L..30C} {730, L30}

\bibitem[\protect\citeauthoryear{Carilli, Gnedin  \& Owen}{Carilli et~al.}{2002}]{Carilli_02_Forest21cm}
Carilli C.~L.,  Gnedin N.~Y.,   Owen F.,  2002, \mn@doi [ApJ] {10.1086/342179}, 577, 22

\bibitem[\protect\citeauthoryear{{Chan}}{{Chan}}{2020}]{Chan2020Thesis}
{Chan} J. Y.~H.,  2020, PhD thesis, UCL (University College London), London, UK

\bibitem[\protect\citeauthoryear{{Chan}, {Wu}, {On}, {Barnes}, {McEwen}  \& {Kitching}}{{Chan} et~al.}{2019}]{Chan2019MNRAS}
{Chan} J. Y.~H.,  {Wu} K.,  {On} A. Y.~L.,  {Barnes} D.~J.,  {McEwen} J.~D.,   {Kitching} T.~D.,  2019, \mn@doi [\mnras] {10.1093/mnras/sty3498}, \href {https://ui.adsabs.harvard.edu/abs/2019MNRAS.484.1427C} {484, 1427}

\bibitem[\protect\citeauthoryear{{Chang}, {Gong}, {Santos}, {Silva}, {Aguirre}, {Dor{\'e}}  \& {Pritchard}}{{Chang} et~al.}{2015}]{Chang2015aska_COCIILya}
{Chang} T.~C.,  {Gong} Y.,  {Santos} M.,  {Silva} M.~B.,  {Aguirre} J.,  {Dor{\'e}} O.,   {Pritchard} J.,  2015, in Advancing Astrophysics with the Square Kilometre Array (AASKA14). p.~4 (\mn@eprint {arXiv} {1501.04654})

\bibitem[\protect\citeauthoryear{{Chen}, {Xu}, {Wang}  \& {Chen}}{{Chen} et~al.}{2019}]{Chen2019ApJ}
{Chen} Z.,  {Xu} Y.,  {Wang} Y.,   {Chen} X.,  2019, \mn@doi [\apj] {10.3847/1538-4357/ab43e6}, \href {https://ui.adsabs.harvard.edu/abs/2019ApJ...885...23C} {885, 23}

\bibitem[\protect\citeauthoryear{{Choudhury}, {Datta}  \& {Majumdar}}{{Choudhury} et~al.}{2022}]{Choudhury2022MNRAS}
{Choudhury} M.,  {Datta} A.,   {Majumdar} S.,  2022, \mn@doi [\mnras] {10.1093/mnras/stac736}, \href {https://ui.adsabs.harvard.edu/abs/2022MNRAS.512.5010C} {512, 5010}

\bibitem[\protect\citeauthoryear{{Ciardi}, {Inoue}, {Mack}, {Xu}  \& {Bernardi}}{{Ciardi} et~al.}{2015}]{Ciardi_15_Forest21SKA}
{Ciardi} B.,  {Inoue} S.,  {Mack} K.,  {Xu} Y.,   {Bernardi} G.,  2015, Proc. Sci., \href {https://ui.adsabs.harvard.edu/abs/2015aska.confE...6C} {p.~6}

\bibitem[\protect\citeauthoryear{{Colless} et~al.,}{{Colless} et~al.}{2003}]{Colless2003astroph_2dFGRS}
{Colless} M.,  et~al., 2003, arXiv e-prints, \href {https://ui.adsabs.harvard.edu/abs/2003astro.ph..6581C} {pp astro--ph/0306581}

\bibitem[\protect\citeauthoryear{{Dayal} et~al.,}{{Dayal} et~al.}{2020}]{Dayal2020MNRAS}
{Dayal} P.,  et~al., 2020, \mn@doi [\mnras] {10.1093/mnras/staa1138}, \href {https://ui.adsabs.harvard.edu/abs/2020MNRAS.495.3065D} {495, 3065}

\bibitem[\protect\citeauthoryear{{DeBoer} et~al.,}{{DeBoer} et~al.}{2017}]{DeBoer2017PASP_HERA}
{DeBoer} D.~R.,  et~al., 2017, \mn@doi [\pasp] {10.1088/1538-3873/129/974/045001}, \href {https://ui.adsabs.harvard.edu/abs/2017PASP..129d5001D} {129, 045001}

\bibitem[\protect\citeauthoryear{{Einstein}}{{Einstein}}{1916}]{Einstein1916}
{Einstein} A.,  1916, Physikalische Gesellschaft Z\&uuml;rich, \href {https://ui.adsabs.harvard.edu/abs/1916PhyGZ..18...47E} {18, 47}

\bibitem[\protect\citeauthoryear{{Einstein}}{{Einstein}}{1917}]{Einstein1917}
{Einstein} A.,  1917, Physikalische Zeitschrift, \href {https://ui.adsabs.harvard.edu/abs/1917PhyZ...18..121E} {18, 121}

\bibitem[\protect\citeauthoryear{Essen, Donaldson, Bangham  \& Hope}{Essen et~al.}{1971}]{Essen71hyperfine}
Essen L.,  Donaldson R.~W.,  Bangham M.~J.,   Hope E.~G.,  1971, Nature, 229, 110

\bibitem[\protect\citeauthoryear{Field}{Field}{1958}]{Field1958IRE_WFeffects}
Field G.~B.,  1958, \mn@doi [Proc. of the IRE] {10.1109/JRPROC.1958.286741}, 46, 240

\bibitem[\protect\citeauthoryear{{Field}}{{Field}}{1959}]{Field1959ApJ_reslineprofile}
{Field} G.~B.,  1959, \mn@doi [\apj] {10.1086/146654}, \href {http://adsabs.harvard.edu/abs/1959ApJ...129..551F} {129, 551}

\bibitem[\protect\citeauthoryear{{Fuerst} \& {Wu}}{{Fuerst} \& {Wu}}{2004}]{Fuerst2004A&A}
{Fuerst} S.~V.,  {Wu} K.,  2004, \mn@doi [\aap] {10.1051/0004-6361:20035814}, \href {https://ui.adsabs.harvard.edu/abs/2004A&A...424..733F} {424, 733}

\bibitem[\protect\citeauthoryear{{Furlanetto}}{{Furlanetto}}{2006}]{Furlanetto_06_Forest21cm}
{Furlanetto} S.~R.,  2006, \mn@doi [\mnras] {10.1111/j.1365-2966.2006.10603.x}, \href {https://ui.adsabs.harvard.edu/abs/2006MNRAS.370.1867F} {370, 1867}

\bibitem[\protect\citeauthoryear{{Furlanetto}}{{Furlanetto}}{2016}]{Furlanetto2016_21cmProbeEoR}
{Furlanetto} S.~R.,  2016, {The 21-cm line as a probe of reionization}.
Springer International Publishing, p.~247, \mn@doi{10.1007/978-3-319-21957-8_9}

\bibitem[\protect\citeauthoryear{{Furlanetto} \& {Lidz}}{{Furlanetto} \& {Lidz}}{2007}]{Furlanetto2007ApJ_21cmCrossCorrGalaxy}
{Furlanetto} S.~R.,  {Lidz} A.,  2007, \mn@doi [\apj] {10.1086/513009}, \href {https://ui.adsabs.harvard.edu/abs/2007ApJ...660.1030F} {660, 1030}

\bibitem[\protect\citeauthoryear{{Furlanetto} \& {Loeb}}{{Furlanetto} \& {Loeb}}{2002}]{FurlanettoLoeb_02_Forest21cm}
{Furlanetto} S.~R.,  {Loeb} A.,  2002, \mn@doi [\apj] {10.1086/342757}, \href {https://ui.adsabs.harvard.edu/abs/2002ApJ...579....1F} {579, 1}

\bibitem[\protect\citeauthoryear{{Furlanetto} \& {Oh}}{{Furlanetto} \& {Oh}}{2005}]{Furlanetto2005MNRAS_TaxingTheRich}
{Furlanetto} S.~R.,  {Oh} S.~P.,  2005, \mn@doi [\mnras] {10.1111/j.1365-2966.2005.09505.x}, \href {http://adsabs.harvard.edu/abs/2005MNRAS.363.1031F} {363, 1031}

\bibitem[\protect\citeauthoryear{{Furlanetto}, {Oh}  \& {Briggs}}{{Furlanetto} et~al.}{2006}]{Furlanetto2006_review}
{Furlanetto} S.~R.,  {Oh} S.~P.,   {Briggs} F.~H.,  2006, \mn@doi [\physrep] {10.1016/j.physrep.2006.08.002}, \href {https://ui.adsabs.harvard.edu/abs/2006PhR...433..181F} {433, 181}

\bibitem[\protect\citeauthoryear{{Geil}, {Mutch}, {Poole}, {Duffy}, {Mesinger}  \& {Wyithe}}{{Geil} et~al.}{2017}]{Geil2017MNRAS}
{Geil} P.~M.,  {Mutch} S.~J.,  {Poole} G.~B.,  {Duffy} A.~R.,  {Mesinger} A.,   {Wyithe} J. S.~B.,  2017, \mn@doi [\mnras] {10.1093/mnras/stx1841}, \href {https://ui.adsabs.harvard.edu/abs/2017MNRAS.472.1324G} {472, 1324}

\bibitem[\protect\citeauthoryear{{Gnedin}}{{Gnedin}}{2000}]{Gnedin2000ApJ}
{Gnedin} N.~Y.,  2000, \mn@doi [\apj] {10.1086/308876}, \href {https://ui.adsabs.harvard.edu/abs/2000ApJ...535..530G} {535, 530}

\bibitem[\protect\citeauthoryear{{Gnedin} \& {Shaver}}{{Gnedin} \& {Shaver}}{2004}]{Gnedin2004ApJ}
{Gnedin} N.~Y.,  {Shaver} P.~A.,  2004, \mn@doi [\apj] {10.1086/420735}, \href {https://ui.adsabs.harvard.edu/abs/2004ApJ...608..611G} {608, 611}

\bibitem[\protect\citeauthoryear{{Gong}, {Cooray}, {Silva}, {Santos}  \& {Lubin}}{{Gong} et~al.}{2011}]{Gong2011ApJ_CO}
{Gong} Y.,  {Cooray} A.,  {Silva} M.~B.,  {Santos} M.~G.,   {Lubin} P.,  2011, \mn@doi [\apjl] {10.1088/2041-8205/728/2/L46}, \href {https://ui.adsabs.harvard.edu/abs/2011ApJ...728L..46G} {728, L46}

\bibitem[\protect\citeauthoryear{{Greig}, {Ting}  \& {Kaurov}}{{Greig} et~al.}{2022}]{Greig2022MNRAS}
{Greig} B.,  {Ting} Y.-S.,   {Kaurov} A.~A.,  2022, \mn@doi [\mnras] {10.1093/mnras/stac977}, \href {https://ui.adsabs.harvard.edu/abs/2022MNRAS.tmp..951G} {}

\bibitem[\protect\citeauthoryear{{Guzzo} et~al.,}{{Guzzo} et~al.}{2014}]{Guzzo2014AA_VIPERS}
{Guzzo} L.,  et~al., 2014, \mn@doi [\aap] {10.1051/0004-6361/201321489}, \href {https://ui.adsabs.harvard.edu/abs/2014A&A...566A.108G} {566, A108}

\bibitem[\protect\citeauthoryear{{HERA Collaboration}}{{HERA Collaboration}}{2022}]{Abdurashidova2022ApJ_HERA}
{HERA Collaboration} 2022, \mn@doi [\apj] {10.3847/1538-4357/ac2ffc}, \href {https://ui.adsabs.harvard.edu/abs/2022ApJ...924...51A} {924, 51}

\bibitem[\protect\citeauthoryear{{Haiman} \& {Holder}}{{Haiman} \& {Holder}}{2003}]{Haiman20162003ApJ}
{Haiman} Z.,  {Holder} G.~P.,  2003, \mn@doi [\apj] {10.1086/377337}, \href {https://ui.adsabs.harvard.edu/abs/2003ApJ...595....1H} {595, 1}

\bibitem[\protect\citeauthoryear{Hellwig, Vessot, Levine, Zitzewitz, Allan  \& Glaze}{Hellwig et~al.}{1970}]{Hellwig1970IEEETIM}
Hellwig H.,  Vessot R.,  Levine M.,  Zitzewitz P.,  Allan D.,   Glaze D.,  1970, \mn@doi [IEEE Trans. Instrum. Meas.] {10.1109/TIM.1970.4313902}, 19, 200

\bibitem[\protect\citeauthoryear{{Hogan} \& {Rees}}{{Hogan} \& {Rees}}{1979}]{HoganRees1979MNRAS_21cm}
{Hogan} C.~J.,  {Rees} M.~J.,  1979, \mn@doi [\mnras] {10.1093/mnras/188.4.791}, \href {https://ui.adsabs.harvard.edu/abs/1979MNRAS.188..791H} {188, 791}

\bibitem[\protect\citeauthoryear{{Huchra} et~al.,}{{Huchra} et~al.}{2012}]{Huchra2012ApJS_2MASSzSurvey}
{Huchra} J.~P.,  et~al., 2012, \mn@doi [\apjs] {10.1088/0067-0049/199/2/26}, \href {https://ui.adsabs.harvard.edu/abs/2012ApJS..199...26H} {199, 26}

\bibitem[\protect\citeauthoryear{{Hummer} \& {Rybicki}}{{Hummer} \& {Rybicki}}{1985}]{Hummer1985ApJ}
{Hummer} D.~G.,  {Rybicki} G.~B.,  1985, \mn@doi [\apj] {10.1086/163232}, \href {https://ui.adsabs.harvard.edu/abs/1985ApJ...293..258H} {293, 258}

\bibitem[\protect\citeauthoryear{{Iliev}, {Mellema}, {Pen}, {Merz}, {Shapiro}  \& {Alvarez}}{{Iliev} et~al.}{2006}]{Iliev2006MNRAS}
{Iliev} I.~T.,  {Mellema} G.,  {Pen} U.~L.,  {Merz} H.,  {Shapiro} P.~R.,   {Alvarez} M.~A.,  2006, \mn@doi [\mnras] {10.1111/j.1365-2966.2006.10502.x}, \href {https://ui.adsabs.harvard.edu/abs/2006MNRAS.369.1625I} {369, 1625}

\bibitem[\protect\citeauthoryear{{Kannan}, {Garaldi}, {Smith}, {Pakmor}, {Springel}, {Vogelsberger}  \& {Hernquist}}{{Kannan} et~al.}{2022}]{Kannan2022MNRAS}
{Kannan} R.,  {Garaldi} E.,  {Smith} A.,  {Pakmor} R.,  {Springel} V.,  {Vogelsberger} M.,   {Hernquist} L.,  2022, \mn@doi [\mnras] {10.1093/mnras/stab3710}, \href {https://ui.adsabs.harvard.edu/abs/2022MNRAS.511.4005K} {511, 4005}

\bibitem[\protect\citeauthoryear{{Koopmans} et~al.,}{{Koopmans} et~al.}{2015}]{Koopmans2015aska}
{Koopmans} L.,  et~al., 2015, in Advancing Astrophysics with the Square Kilometre Array (AASKA14). p.~1 (\mn@eprint {arXiv} {1505.07568}), \mn@doi{10.22323/1.215.0001}

\bibitem[\protect\citeauthoryear{{Koopmans} et~al.,}{{Koopmans} et~al.}{2021}]{Koopmans2019ExA}
{Koopmans} L.,  et~al., 2021, \mn@doi [Exp. Astron.] {10.1007/s10686-021-09743-7}, \href {https://ui.adsabs.harvard.edu/abs/2019arXiv190804296K} {pp 1641--1676}

\bibitem[\protect\citeauthoryear{{Kulkarni}, {Choudhury}, {Puchwein}  \& {Haehnelt}}{{Kulkarni} et~al.}{2017}]{Kulkarni2017MNRAS_21cmAGNEoR}
{Kulkarni} G.,  {Choudhury} T.~R.,  {Puchwein} E.,   {Haehnelt} M.~G.,  2017, \mn@doi [\mnras] {10.1093/mnras/stx1167}, \href {https://ui.adsabs.harvard.edu/abs/2017MNRAS.469.4283K} {469, 4283}

\bibitem[\protect\citeauthoryear{{Libeskind} et~al.,}{{Libeskind} et~al.}{2018}]{Libeskind2018MNRAS}
{Libeskind} N.~I.,  et~al., 2018, \mn@doi [\mnras] {10.1093/mnras/stx1976}, \href {https://ui.adsabs.harvard.edu/abs/2018MNRAS.473.1195L} {473, 1195}

\bibitem[\protect\citeauthoryear{{Lidz}, {Zahn}, {Furlanetto}, {McQuinn}, {Hernquist}  \& {Zaldarriaga}}{{Lidz} et~al.}{2009}]{Lidz2009ApJ}
{Lidz} A.,  {Zahn} O.,  {Furlanetto} S.~R.,  {McQuinn} M.,  {Hernquist} L.,   {Zaldarriaga} M.,  2009, \mn@doi [\apj] {10.1088/0004-637X/690/1/252}, \href {https://ui.adsabs.harvard.edu/abs/2009ApJ...690..252L} {690, 252}

\bibitem[\protect\citeauthoryear{{Lidz}, {Furlanetto}, {Oh}, {Aguirre}, {Chang}, {Dor{\'e}}  \& {Pritchard}}{{Lidz} et~al.}{2011}]{Lidz2011ApJ_21cmCrossCorrCO}
{Lidz} A.,  {Furlanetto} S.~R.,  {Oh} S.~P.,  {Aguirre} J.,  {Chang} T.-C.,  {Dor{\'e}} O.,   {Pritchard} J.~R.,  2011, \mn@doi [\apj] {10.1088/0004-637X/741/2/70}, \href {https://ui.adsabs.harvard.edu/abs/2011ApJ...741...70L} {741, 70}

\bibitem[\protect\citeauthoryear{{Loeb} \& {Furlanetto}}{{Loeb} \& {Furlanetto}}{2013}]{LoebFurlanetto2013Book}
{Loeb} A.,  {Furlanetto} S.~R.,  2013, {The first galaxies in the Universe}.
Princeton University Press, Princeton, NJ

\bibitem[\protect\citeauthoryear{{Mack} \& {Wyithe}}{{Mack} \& {Wyithe}}{2012}]{Mack_12_Forest21cmDetect}
{Mack} K.~J.,  {Wyithe} J.~S.~B.,  2012, \mn@doi [\mnras] {10.1111/j.1365-2966.2012.21561.x}, \href {https://ui.adsabs.harvard.edu/abs/2012MNRAS.425.2988M} {425, 2988}

\bibitem[\protect\citeauthoryear{{Madau}, {Meiksin}  \& {Rees}}{{Madau} et~al.}{1997}]{Madau1997ApJ}
{Madau} P.,  {Meiksin} A.,   {Rees} M.~J.,  1997, \mn@doi [\apj] {10.1086/303549}, \href {http://adsabs.harvard.edu/abs/1997ApJ...475..429M} {475, 429}

\bibitem[\protect\citeauthoryear{{Majumdar}, {Pritchard}, {Mondal}, {Watkinson}, {Bharadwaj}  \& {Mellema}}{{Majumdar} et~al.}{2018}]{Suman2018MNRAS}
{Majumdar} S.,  {Pritchard} J.~R.,  {Mondal} R.,  {Watkinson} C.~A.,  {Bharadwaj} S.,   {Mellema} G.,  2018, \mn@doi [\mnras] {10.1093/mnras/sty535}, \href {https://ui.adsabs.harvard.edu/abs/2018MNRAS.476.4007M} {476, 4007}

\bibitem[\protect\citeauthoryear{{Mather} et~al.,}{{Mather} et~al.}{1994}]{Mather1994ApJ}
{Mather} J.~C.,  et~al., 1994, \mn@doi [\apj] {10.1086/173574}, \href {http://adsabs.harvard.edu/abs/1994ApJ...420..439M} {420, 439}

\bibitem[\protect\citeauthoryear{{McQuinn}, {Lidz}, {Zahn}, {Dutta}, {Hernquist}  \& {Zaldarriaga}}{{McQuinn} et~al.}{2007}]{McQuinn2007MNRAS}
{McQuinn} M.,  {Lidz} A.,  {Zahn} O.,  {Dutta} S.,  {Hernquist} L.,   {Zaldarriaga} M.,  2007, \mn@doi [\mnras] {10.1111/j.1365-2966.2007.11489.x}, \href {https://ui.adsabs.harvard.edu/abs/2007MNRAS.377.1043M} {377, 1043}

\bibitem[\protect\citeauthoryear{{Mellema}, {Koopmans}, {Shukla}, {Datta}, {Mesinger}  \& {Majumdar}}{{Mellema} et~al.}{2015}]{Mellema2015aaska}
{Mellema} G.,  {Koopmans} L.,  {Shukla} H.,  {Datta} K.~K.,  {Mesinger} A.,   {Majumdar} S.,  2015, in Advancing Astrophysics with the Square Kilometre Array (AASKA14). p.~10 (\mn@eprint {arXiv} {1501.04203})

\bibitem[\protect\citeauthoryear{{Mesinger}}{{Mesinger}}{2016}]{Mesinger2016Book}
{Mesinger} A.,  2016, {Understanding the Epoch of cosmic Reionization: challenges and progress}.
 Astrophysics and space science library Vol. 423, Springer International Publishing, \mn@doi{10.1007/978-3-319-21957-8}

\bibitem[\protect\citeauthoryear{{Mesinger}, {Furlanetto}  \& {Cen}}{{Mesinger} et~al.}{2011}]{Mesinger2011MNRAS}
{Mesinger} A.,  {Furlanetto} S.,   {Cen} R.,  2011, \mn@doi [\mnras] {10.1111/j.1365-2966.2010.17731.x}, \href {https://ui.adsabs.harvard.edu/abs/2011MNRAS.411..955M} {411, 955}

\bibitem[\protect\citeauthoryear{{Mihalas}}{{Mihalas}}{1978}]{Mihalas1978book}
{Mihalas} D.,  1978, {Stellar atmospheres}.
W H Freeman \& Co

\bibitem[\protect\citeauthoryear{{Mohankumar} \& {Sen}}{{Mohankumar} \& {Sen}}{2019}]{Mohankumar2019VoigtComputation}
{Mohankumar} N.,  {Sen} S.,  2019, \mn@doi [\jqsrt] {10.1016/j.jqsrt.2018.11.022}, \href {https://ui.adsabs.harvard.edu/abs/2019JQSRT.224..192M} {224, 192}

\bibitem[\protect\citeauthoryear{{Morales} \& {Wyithe}}{{Morales} \& {Wyithe}}{2010}]{Morale2010ARAA}
{Morales} M.~F.,  {Wyithe} J. S.~B.,  2010, \mn@doi [\araa] {10.1146/annurev-astro-081309-130936}, \href {https://ui.adsabs.harvard.edu/abs/2010ARA&A..48..127M} {48, 127}

\bibitem[\protect\citeauthoryear{{Mu{\~n}oz}, {Qin}, {Mesinger}, {Murray}, {Greig}  \& {Mason}}{{Mu{\~n}oz} et~al.}{2022}]{Munoz2022MNRAS}
{Mu{\~n}oz} J.~B.,  {Qin} Y.,  {Mesinger} A.,  {Murray} S.~G.,  {Greig} B.,   {Mason} C.,  2022, \mn@doi [\mnras] {10.1093/mnras/stac185}, \href {https://ui.adsabs.harvard.edu/abs/2022MNRAS.511.3657M} {511, 3657}

\bibitem[\protect\citeauthoryear{{Nusser}}{{Nusser}}{2005}]{Adi2005MNRAS}
{Nusser} A.,  2005, \mn@doi [\mnras] {10.1111/j.1365-2966.2005.08894.x}, \href {https://ui.adsabs.harvard.edu/abs/2005MNRAS.359..183N} {359, 183}

\bibitem[\protect\citeauthoryear{Peacock}{Peacock}{1999}]{Peacock1999Book}
Peacock J.~A.,  1999, Cosmological physics.
Cambridge astrophysics, Cambridge University Press, Cambridge, \url {https://books.google.co.uk/books?id=t8O-yylU0j0C}

\bibitem[\protect\citeauthoryear{{Penzias} \& {Wilson}}{{Penzias} \& {Wilson}}{1965}]{WilsonPenzias1965ApJ}
{Penzias} A.~A.,  {Wilson} R.~W.,  1965, \mn@doi [\apj] {10.1086/148307}, \href {http://adsabs.harvard.edu/abs/1965ApJ...142..419P} {142, 419}

\bibitem[\protect\citeauthoryear{{Philip} et~al.,}{{Philip} et~al.}{2019}]{Philp2019JAI_PRIZM}
{Philip} L.,  et~al., 2019, \mn@doi [Journal of Astronomical Instrumentation] {10.1142/S2251171719500041}, \href {https://ui.adsabs.harvard.edu/abs/2019JAI.....850004P} {8, 1950004}

\bibitem[\protect\citeauthoryear{{Planck Collaboration XIII}}{{Planck Collaboration XIII}}{2016}]{Planck2015_CosmoPara}
{Planck Collaboration XIII} 2016, \mn@doi [A\&A] {10.1051/0004-6361/201525830}, 594, A13

\bibitem[\protect\citeauthoryear{{Planck Collaboration XVI}}{{Planck Collaboration XVI}}{2014}]{Planck13}
{Planck Collaboration XVI} 2014, \mn@doi [\aap] {10.1051/0004-6361/201321591}, \href {https://ui.adsabs.harvard.edu/abs/2014A&A...571A..16P} {571, A16}

\bibitem[\protect\citeauthoryear{{Price} et~al.,}{{Price} et~al.}{2018}]{Price2018MNRAS_LEDA}
{Price} D.~C.,  et~al., 2018, \mn@doi [\mnras] {10.1093/mnras/sty1244}, \href {https://ui.adsabs.harvard.edu/abs/2018MNRAS.478.4193P} {478, 4193}

\bibitem[\protect\citeauthoryear{{Pritchard} \& {Loeb}}{{Pritchard} \& {Loeb}}{2012}]{Pritchard2012Review}
{Pritchard} J.~R.,  {Loeb} A.,  2012, \mn@doi [Rep. Prog. Phys.] {10.1088/0034-4885/75/8/086901}, \href {http://adsabs.harvard.edu/abs/2012RPPh...75h6901P} {75, 086901}

\bibitem[\protect\citeauthoryear{{Rien van de Weygaert}, {Sergei Shandarin}, {Enn Saar}  \& {Jaan Einasto}}{{Rien van de Weygaert} et~al.}{2016}]{vandeWeygaert2016IAUS}
{Rien van de Weygaert} {Sergei Shandarin} {Enn Saar}  {Jaan Einasto} eds, 2016, The Zeldovich Universe: Genesis and Growth of the Cosmic Web ~ Vol. 308, \mn@doi{10.1017/S174392131601098X.
}

\bibitem[\protect\citeauthoryear{{Robertson}, {Ellis}, {Dunlop}, {McLure}  \& {Stark}}{{Robertson} et~al.}{2010}]{Robertson2010Nature}
{Robertson} B.~E.,  {Ellis} R.~S.,  {Dunlop} J.~S.,  {McLure} R.~J.,   {Stark} D.~P.,  2010, \mn@doi [\nat] {10.1038/nature09527}, \href {https://ui.adsabs.harvard.edu/abs/2010Natur.468...49R} {468, 49}

\bibitem[\protect\citeauthoryear{{Rutten}}{{Rutten}}{2003}]{Rutten2003RTNotes}
{Rutten} R.~J.,  2003, {Radiative transfer in stellar atmospheres}.
Sterrekundig Instuut Utrecht

\bibitem[\protect\citeauthoryear{{Rybicki} \& {Hummer}}{{Rybicki} \& {Hummer}}{1978}]{Rybicki1978ApJ}
{Rybicki} G.~B.,  {Hummer} D.~G.,  1978, \mn@doi [\apj] {10.1086/155826}, \href {https://ui.adsabs.harvard.edu/abs/1978ApJ...219..654R} {219, 654}

\bibitem[\protect\citeauthoryear{{Santos}, {Amblard}, {Pritchard}, {Trac}, {Cen}  \& {Cooray}}{{Santos} et~al.}{2008}]{Santos2008ApJ}
{Santos} M.~G.,  {Amblard} A.,  {Pritchard} J.,  {Trac} H.,  {Cen} R.,   {Cooray} A.,  2008, \mn@doi [\apj] {10.1086/592487}, \href {https://ui.adsabs.harvard.edu/abs/2008ApJ...689....1S} {689, 1}

\bibitem[\protect\citeauthoryear{{Santos}, {Ferramacho}, {Silva}, {Amblard}  \& {Cooray}}{{Santos} et~al.}{2010}]{Santos2010MNRAS}
{Santos} M.~G.,  {Ferramacho} L.,  {Silva} M.~B.,  {Amblard} A.,   {Cooray} A.,  2010, \mn@doi [\mnras] {10.1111/j.1365-2966.2010.16898.x}, \href {https://ui.adsabs.harvard.edu/abs/2010MNRAS.406.2421S} {406, 2421}

\bibitem[\protect\citeauthoryear{{Schreier}}{{Schreier}}{1992}]{Schreier1992VoigtComputation}
{Schreier} F.,  1992, \mn@doi [\jqsrt] {10.1016/0022-4073(92)90139-U}, \href {https://ui.adsabs.harvard.edu/abs/1992JQSRT..48..743S} {48, 743}

\bibitem[\protect\citeauthoryear{{Scott} \& {Rees}}{{Scott} \& {Rees}}{1990}]{Scott1990MNRAS}
{Scott} D.,  {Rees} M.~J.,  1990, \mnras, \href {https://ui.adsabs.harvard.edu/abs/1990MNRAS.247..510S} {247, 510}

\bibitem[\protect\citeauthoryear{{Shin}, {Trac}  \& {Cen}}{{Shin} et~al.}{2008}]{Shin2008ApJ}
{Shin} M.-S.,  {Trac} H.,   {Cen} R.,  2008, \mn@doi [\apj] {10.1086/588247}, \href {https://ui.adsabs.harvard.edu/abs/2008ApJ...681..756S} {681, 756}

\bibitem[\protect\citeauthoryear{{Silva}, {Santos}, {Gong}, {Cooray}  \& {Bock}}{{Silva} et~al.}{2013}]{Silva2013ApJ_LyAlphaCrossCorr21cm}
{Silva} M.~B.,  {Santos} M.~G.,  {Gong} Y.,  {Cooray} A.,   {Bock} J.,  2013, \mn@doi [\apj] {10.1088/0004-637X/763/2/132}, \href {https://ui.adsabs.harvard.edu/abs/2013ApJ...763..132S} {763, 132}

\bibitem[\protect\citeauthoryear{{Silva} et~al.,}{{Silva} et~al.}{2021}]{Silva2021ExA}
{Silva} M.~B.,  et~al., 2021, \mn@doi [Experimental Astronomy] {10.1007/s10686-021-09755-3}, \href {https://ui.adsabs.harvard.edu/abs/2021ExA....51.1593S} {51, 1593}

\bibitem[\protect\citeauthoryear{{Singh}, {Subrahmanyan}, {Shankar}, {Rao}, {Girish}, {Raghunathan}, {Somashekar}  \& {Srivani}}{{Singh} et~al.}{2018}]{Singh2018ExA_SARA2}
{Singh} S.,  {Subrahmanyan} R.,  {Shankar} N.~U.,  {Rao} M.~S.,  {Girish} B.~S.,  {Raghunathan} A.,  {Somashekar} R.,   {Srivani} K.~S.,  2018, \mn@doi [Experimental Astronomy] {10.1007/s10686-018-9584-3}, \href {https://ui.adsabs.harvard.edu/abs/2018ExA....45..269S} {45, 269}

\bibitem[\protect\citeauthoryear{{Smoot} et~al.,}{{Smoot} et~al.}{1992}]{Smoot92}
{Smoot} G.~F.,  et~al., 1992, \mn@doi [\apjl] {10.1086/186504}, \href {http://adsabs.harvard.edu/abs/1992ApJ...396L...1S} {396, L1}

\bibitem[\protect\citeauthoryear{Spergel et~al.,}{Spergel et~al.}{2003}]{Spergel2003ApJS}
Spergel D.~N.,  et~al., 2003, ApJS, 148, 175

\bibitem[\protect\citeauthoryear{{Struble} \& {Rood}}{{Struble} \& {Rood}}{1999}]{Struble1999GCvelodisp}
{Struble} M.~F.,  {Rood} H.~J.,  1999, \mn@doi [\apjs] {10.1086/313274}, \href {https://ui.adsabs.harvard.edu/abs/1999ApJS..125...35S} {125, 35}

\bibitem[\protect\citeauthoryear{{Subramanian} \& {Padmanabhan}}{{Subramanian} \& {Padmanabhan}}{1993}]{Subramanian1993MNRAS}
{Subramanian} K.,  {Padmanabhan} T.,  1993, \mn@doi [\mnras] {10.1093/mnras/265.1.101}, \href {https://ui.adsabs.harvard.edu/abs/1993MNRAS.265..101S} {265, 101}

\bibitem[\protect\citeauthoryear{{Sunyaev} \& {Zeldovich}}{{Sunyaev} \& {Zeldovich}}{1975}]{SunyaevZeldovich1975MNRAS_21cmGalaxiesBirth}
{Sunyaev} R.~A.,  {Zeldovich} I.~B.,  1975, \mn@doi [\mnras] {10.1093/mnras/171.2.375}, \href {https://ui.adsabs.harvard.edu/abs/1975MNRAS.171..375S} {171, 375}

\bibitem[\protect\citeauthoryear{{Tegmark} et~al.,}{{Tegmark} et~al.}{2004}]{Tegmark2004ApJ_SLOAN}
{Tegmark} M.,  et~al., 2004, \mn@doi [\apj] {10.1086/382125}, \href {https://ui.adsabs.harvard.edu/abs/2004ApJ...606..702T} {606, 702}

\bibitem[\protect\citeauthoryear{{Tingay} et~al.,}{{Tingay} et~al.}{2013}]{Tingay2013PASA_MWA}
{Tingay} S.~J.,  et~al., 2013, \mn@doi [\pasa] {10.1017/pasa.2012.007}, \href {https://ui.adsabs.harvard.edu/abs/2013PASA...30....7T} {30, e007}

\bibitem[\protect\citeauthoryear{{Tucker}}{{Tucker}}{1977}]{Tucker1977Book}
{Tucker} H.~W.,  1977, {Radiation processes in astrophysics}.
The MIT Press - Cambridge, MA, London

\bibitem[\protect\citeauthoryear{{Vernstrom}, {Gaensler}, {Vacca}, {Farnes}, {Haverkorn}  \& {O'Sullivan}}{{Vernstrom} et~al.}{2018}]{Vernstrom2018QuasarRadio}
{Vernstrom} T.,  {Gaensler} B.~M.,  {Vacca} V.,  {Farnes} J.~S.,  {Haverkorn} M.,   {O'Sullivan} S.~P.,  2018, \mn@doi [\mnras] {10.1093/mnras/stx3191}, \href {https://ui.adsabs.harvard.edu/abs/2018MNRAS.475.1736V} {475, 1736}

\bibitem[\protect\citeauthoryear{{Visbal} \& {Loeb}}{{Visbal} \& {Loeb}}{2010}]{Visbal2010JCAP}
{Visbal} E.,  {Loeb} A.,  2010, \mn@doi [\jcap] {10.1088/1475-7516/2010/11/016}, \href {https://ui.adsabs.harvard.edu/abs/2010JCAP...11..016V} {2010, 016}

\bibitem[\protect\citeauthoryear{{Voytek}, {Natarajan}, {J{\'a}uregui Garc{\'\i}a}, {Peterson}  \& {L{\'o}pez-Cruz}}{{Voytek} et~al.}{2014}]{Voytek2014ApJ_SCIHI}
{Voytek} T.~C.,  {Natarajan} A.,  {J{\'a}uregui Garc{\'\i}a} J.~M.,  {Peterson} J.~B.,   {L{\'o}pez-Cruz} O.,  2014, \mn@doi [\apjl] {10.1088/2041-8205/782/1/L9}, \href {https://ui.adsabs.harvard.edu/abs/2014ApJ...782L...9V} {782, L9}

\bibitem[\protect\citeauthoryear{{Watkinson} \& {Pritchard}}{{Watkinson} \& {Pritchard}}{2014}]{Watkinson2014MNRAS}
{Watkinson} C.~A.,  {Pritchard} J.~R.,  2014, \mn@doi [\mnras] {10.1093/mnras/stu1384}, \href {https://ui.adsabs.harvard.edu/abs/2014MNRAS.443.3090W} {443, 3090}

\bibitem[\protect\citeauthoryear{{Watkinson}, {Majumdar}, {Pritchard}  \& {Mondal}}{{Watkinson} et~al.}{2017}]{Watkinson2017MNRAS}
{Watkinson} C.~A.,  {Majumdar} S.,  {Pritchard} J.~R.,   {Mondal} R.,  2017, \mn@doi [\mnras] {10.1093/mnras/stx2130}, \href {https://ui.adsabs.harvard.edu/abs/2017MNRAS.472.2436W} {472, 2436}

\bibitem[\protect\citeauthoryear{{Watkinson}, {Giri}, {Ross}, {Dixon}, {Iliev}, {Mellema}  \& {Pritchard}}{{Watkinson} et~al.}{2019}]{Watkinson2019MNRAS}
{Watkinson} C.~A.,  {Giri} S.~K.,  {Ross} H.~E.,  {Dixon} K.~L.,  {Iliev} I.~T.,  {Mellema} G.,   {Pritchard} J.~R.,  2019, \mn@doi [\mnras] {10.1093/mnras/sty2740}, \href {https://ui.adsabs.harvard.edu/abs/2019MNRAS.482.2653W} {482, 2653}

\bibitem[\protect\citeauthoryear{{White}, {Frenk}, {Davis}  \& {Efstathiou}}{{White} et~al.}{1987}]{White1987ApJ_Nbody}
{White} S. D.~M.,  {Frenk} C.~S.,  {Davis} M.,   {Efstathiou} G.,  1987, \mn@doi [\apj] {10.1086/164990}, \href {https://ui.adsabs.harvard.edu/abs/1987ApJ...313..505W} {313, 505}

\bibitem[\protect\citeauthoryear{{Willott}, {Rawlings}, {Blundell}  \& {Lacy}}{{Willott} et~al.}{1998}]{Willott1998RadioLoudQSOs}
{Willott} C.~J.,  {Rawlings} S.,  {Blundell} K.~M.,   {Lacy} M.,  1998, \mn@doi [\mnras] {10.1046/j.1365-8711.1998.01946.x}, \href {https://ui.adsabs.harvard.edu/abs/1998MNRAS.300..625W} {300, 625}

\bibitem[\protect\citeauthoryear{{Wouthuysen}}{{Wouthuysen}}{1952}]{Wouthuysen1952AJ}
{Wouthuysen} S.~A.,  1952, \mn@doi [\aj] {10.1086/106661}, \href {https://ui.adsabs.harvard.edu/abs/1952AJ.....57R..31W} {57, 31}

\bibitem[\protect\citeauthoryear{{Wu}, {Soria}, {Hunstead}  \& {Johnston}}{{Wu} et~al.}{2001}]{Wu2001MNRAS}
{Wu} K.,  {Soria} R.,  {Hunstead} R.~W.,   {Johnston} H.~M.,  2001, \mn@doi [\mnras] {10.1046/j.1365-8711.2001.03915.x}, \href {https://ui.adsabs.harvard.edu/abs/2001MNRAS.320..177W} {320, 177}

\bibitem[\protect\citeauthoryear{{Wu}, {Han}  \& {Chan}}{{Wu} et~al.}{2023}]{Wu2023MNRAS}
{Wu} K.,  {Han} Q.,   {Chan} J. Y.~H.,  2023, \mnras, submitted

\bibitem[\protect\citeauthoryear{{Wyithe} \& {Loeb}}{{Wyithe} \& {Loeb}}{2008}]{WyitheLoeb2008MNRAS}
{Wyithe} J. S.~B.,  {Loeb} A.,  2008, \mn@doi [\mnras] {10.1111/j.1365-2966.2007.12568.x}, \href {https://ui.adsabs.harvard.edu/abs/2008MNRAS.383..606W} {383, 606}

\bibitem[\protect\citeauthoryear{{Xu}, {Chen}, {Fan}, {Trac}  \& {Cen}}{{Xu} et~al.}{2009}]{Xu_09_Forest21IGMTemp}
{Xu} Y.,  {Chen} X.,  {Fan} Z.,  {Trac} H.,   {Cen} R.,  2009, \mn@doi [\apj] {10.1088/0004-637X/704/2/1396}, \href {https://ui.adsabs.harvard.edu/abs/2009ApJ...704.1396X} {704, 1396}

\bibitem[\protect\citeauthoryear{{Younsi}, {Wu}  \& {Fuerst}}{{Younsi} et~al.}{2012}]{Younsi2012AA_GRRT}
{Younsi} Z.,  {Wu} K.,   {Fuerst} S.~V.,  2012, \mn@doi [\aap] {10.1051/0004-6361/201219599}, \href {https://ui.adsabs.harvard.edu/abs/2012A&A...545A..13Y} {545, A13}

\bibitem[\protect\citeauthoryear{{Zaldarriaga}, {Furlanetto}  \& {Hernquist}}{{Zaldarriaga} et~al.}{2004}]{Zaldarriaga2004ApJ}
{Zaldarriaga} M.,  {Furlanetto} S.~R.,   {Hernquist} L.,  2004, \mn@doi [\apj] {10.1086/386327}, \href {https://ui.adsabs.harvard.edu/abs/2004ApJ...608..622Z} {608, 622}

\bibitem[\protect\citeauthoryear{{de Lapparent}, {Geller}  \& {Huchra}}{{de Lapparent} et~al.}{1986}]{deLapparent1986ApJ}
{de Lapparent} V.,  {Geller} M.~J.,   {Huchra} J.~P.,  1986, \mn@doi [\apjl] {10.1086/184625}, \href {https://ui.adsabs.harvard.edu/abs/1986ApJ...302L...1D} {302, L1}

\bibitem[\protect\citeauthoryear{{van Haarlem} et~al.,}{{van Haarlem} et~al.}{2013}]{vanHaarlem2013AA_LOFAR}
{van Haarlem} M.~P.,  et~al., 2013, \mn@doi [\aap] {10.1051/0004-6361/201220873}, \href {https://ui.adsabs.harvard.edu/abs/2013A&A...556A...2V} {556, A2}

\bibitem[\protect\citeauthoryear{{van de Weygaert} \& {Schaap}}{{van de Weygaert} \& {Schaap}}{2009}]{vandeWeygaert2009LNP_2dFGRS}
{van de Weygaert} R.,  {Schaap} W.,  2009, {The Cosmic Web: Geometric Analysis}.
Springer, pp 291--413, \mn@doi{10.1007/978-3-540-44767-2\_11}

\makeatother
\end{thebibliography}



\appendix


\section{Code convergence and Optimisation} 
\label{app:C21LRTAlgo}

\subsection{Optimisation of redshift and frequency grids}
\label{subsec:Optimisation} 

In logarithmic space, 
  \eqn{\ref{eq:cosmologicalFreqShift}} 
  for the ray-tracing becomes   
\begin{align}
  \log{\nu_{z'}} - \log{\nu_{z}}  & = \log{(1+z')} - \log{(1+z)} \  . 
\label{eq:logFreqlog1plusz}
\end{align}  
Consider the intervals    
    $\Delta_{\log{\nu}} \equiv \log{\nu_{z'}}-\log{\nu_{z}}$, 
    and   
    $\Delta_{\log{(1+z)}} \equiv \log{(1+z')} - \log{(1+z)}$.  
Then, 
  \eqn{\ref{eq:logFreqlog1plusz}} may be expressed as    
\begin{align}
    \Delta_{\log{\nu}} = \Delta_{\log{(1+z)}} \ . 
\end{align} 

\begin{figure}
     \centering
     \includegraphics[width=0.76\columnwidth, trim = 0.0cm 0.0cm 0.cm 0.cm, clip=true]{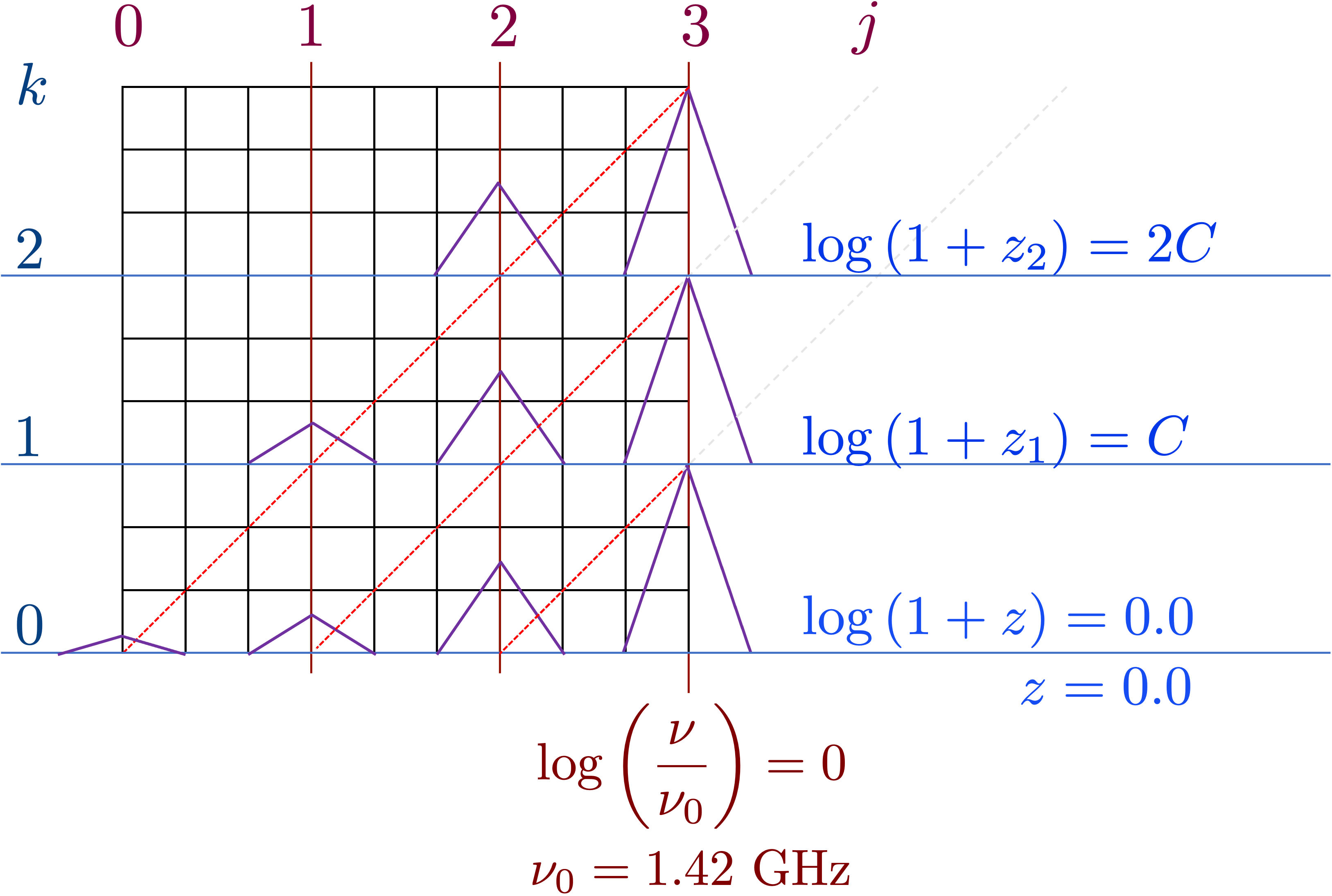}
      \caption[Illustration of the C21LRT square computational grid]
      {An 
      illustration of a two-dimensional C21LRT square computational grid. 
      The grid runs through the indices ``$k$'' (for redshift) 
      and ``$j$'' (for frequency). 
      Equal uniform samplings 
      in $\log(1+z)$, through the index ``$k$'', 
      and in $\log{\nu}$, through the index ``$j$''
      are adopted. 
      The propagation of a ray 
 is a diagonal shift in the computational grid coordinates 
      $(j,k)$. 
 The C21LRT calculations are performed along the rays 
  as those indicated by the red dotted lines. 
      } 
      \label{fig:21cmSquaregrid} 
\end{figure}

\begin{figure}
     \centering
\subfloat{
\begin{minipage}[c]{1.0\linewidth}
     \includegraphics[width=1.0\textwidth, trim = 0.0cm 0.0cm 0.cm 0.cm, clip=true]{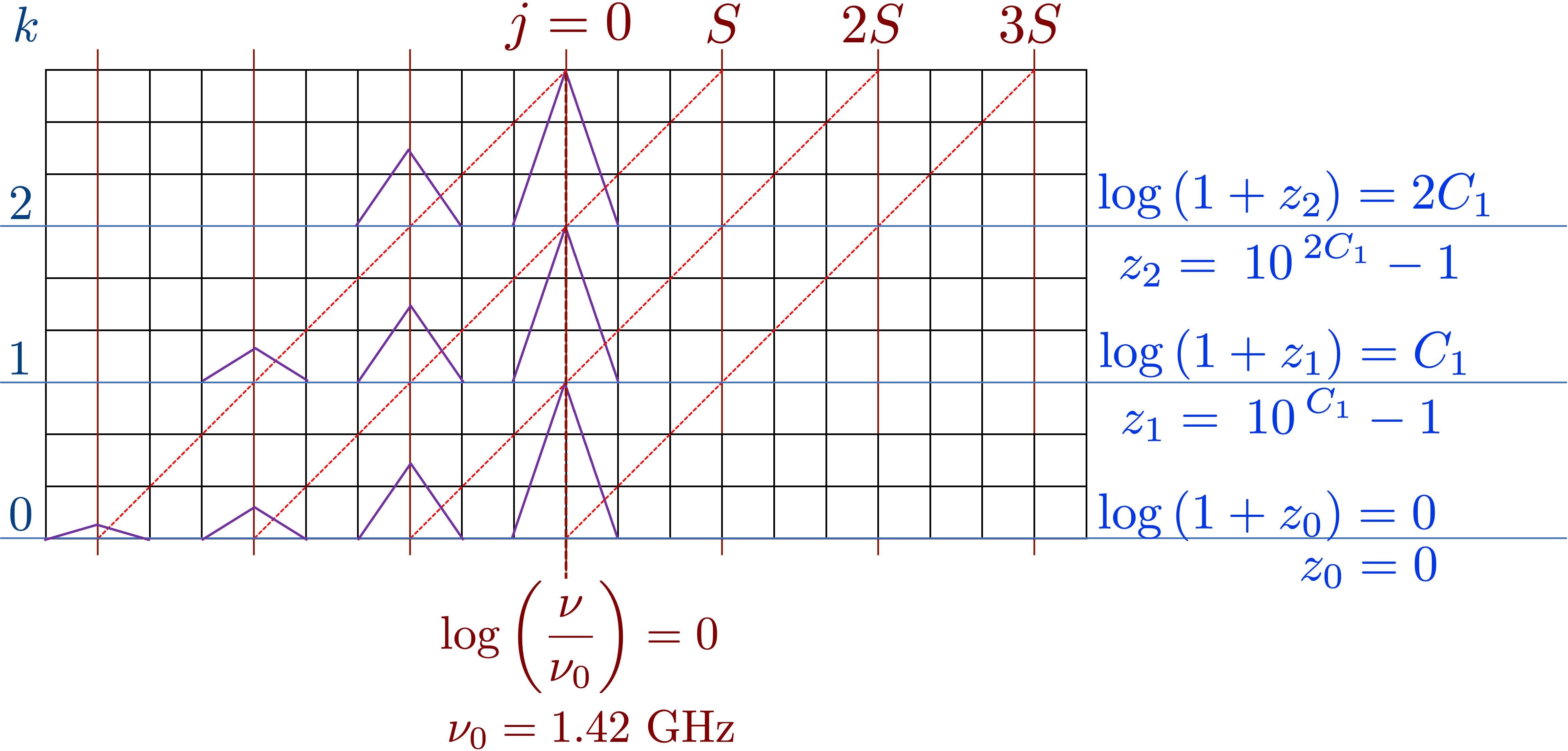}
     \label{fig:21cmRectGrid} 
     \end{minipage}
     } 
\vspace*{-0.25cm}
\subfloat{
\begin{minipage}[c]{1.0\linewidth}
     \includegraphics[width=1.0\textwidth,  trim = 0.0cm 0.0cm 0.cm 0.cm, clip=true]{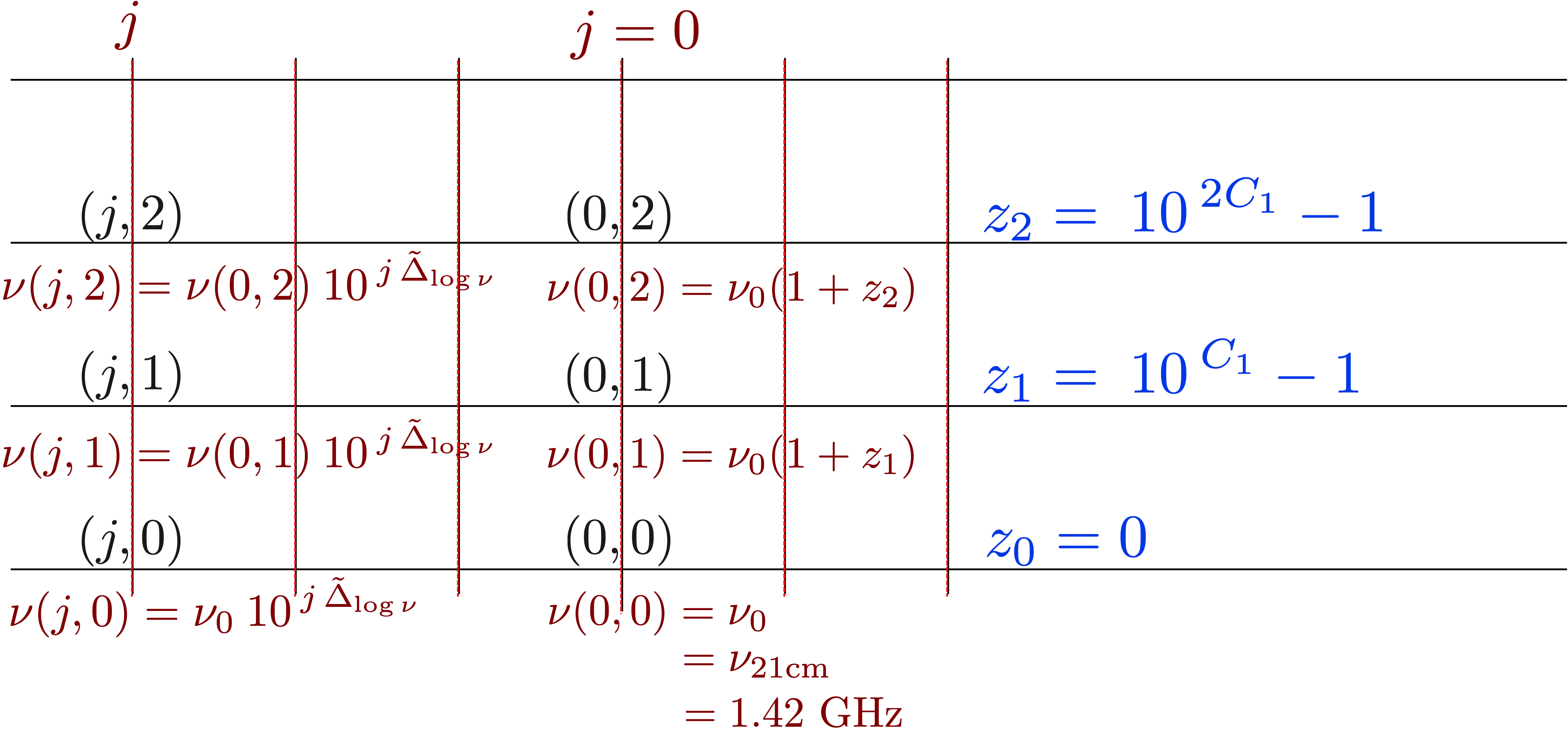} 
     \label{fig:21cmFreqZGrid}
\end{minipage}
} 
\vspace*{-0.25cm}
\caption[Illustration of the C21LRT rectangular computational grid]{
      An illustration of the two-dimensional C21LRT rectangular computational grid (top panel),   
      where frequency can be ``reallocated'' to each grid point 
      (shown in the bottom panel). 
      A uniform sampling is adopted in 
      both $\log(1+z)$ and $\log{\nu}$. 
      The propagation of rays over cosmological redshift,  
      represented by the red dotted lines in the top panel,   
      corresponds to 
      the tracing of the rays through the index ``$k$'' 
      with fixed ``$j$'' in the lattice in the bottom panel.} 
\label{fig:C21LRTRectGridRayTrace}
\end{figure}

The computational grid is specified 
  by the coordinates $(j,k)$,  
  where the index ``$k$'' runs  
  through the redshift and 
  the index ``$j$'' through 
  the radiation frequency.  
For a uniform sampling  
   in $\log(1+z)$, through the index ``$k$'', 
   and in $\log{\nu}$, through the index ``$j$'', 
   $\Delta_{\log{\nu}} = \Delta_{\log{(1+z)}} = C$, 
   where $C$ is a positive constant. 
This gives a $(j,k)$ lattice,   
  and the ray tracing over cosmic time with a diagonal shift in the lattice 
      (see \fig{\ref{fig:21cmSquaregrid}}). 
The sampling in $\Delta_{\log{\nu}} 
   = \Delta_{\log{(1+z)}} = C$ 
  across the redshift  
  is implemented such that 
  the line is sufficiently resolved 
  to reveal the relevant physics of interest. 
Generally, 
   the physical conditions evolve 
   over the redshift on a slower rate 
   than the line profile  
   along the ray.  
For an appropriately chosen sampling in the frequency, 
  a less dense sampling in the redshift 
  will be sufficient in most of the situations.  
An optimal scheme with $\Delta_{\log{(1+z)}} = C_{1}$  
   in the redshift sampling   
    and $\Delta_{\log{\nu}} = C_{2}$   
    in the frequency sampling, 
    where $C_1$ and $C_2$ are positive constants, 
    and their ratio $C_1/C_{2}$ 
    is set to be a fixed positive integer $S$, 
    is adopted in the ray-tracing for the C21LRT calculations.   
The rectangular grid for the ray-tracing 
  is illustrated in 
\fig{\ref{fig:C21LRTRectGridRayTrace}}.   

The ray is traced along the same $j$ over a descending $k$ 
  (from a high redshift $z_{k}$ to $z_0=0$ at $k=0$).  
Along the ray, 
  each grid point $(j, k)$ 
  has a uniquely assigned redshift, 
  given by $z_{k} = 10^{k\,\Delta_{\log{1+z}}}-1$, 
  where $\Delta_{\log{(1+z)}} = C_1 
  = [\log{(1+z_{\rm max})}-\log{(1+z_{0}})]/N_z$. 
Also, a specific frequency is assigned to it, 
  satisfying the rectangular grid specification 
  as described above. 
At $k=0$, 
   $z_{0} = z_{\rm obs} = 0$, 
   and $\nu(j,0)$ is specified 
  by the frequency from $\nu_{\rm max}$ to $\nu_{\rm min}$,  
  with the interval (resolution) 
  $\tilde{\Delta}_{\log{(\nu)}}\big|_{z_{0}} 
    = [\log{\nu_{\rm max}}-\log{\nu_{\rm min}})/N_{j}\big|_{z_{0}} = C_{2}$. 
The radiation frequencies at higher redshifts 
  are assigned by $\nu(j,k)|_{z_{k}} = \nu(j,0) \times (1+z_{k})$, 
  satisfying \eqn{\ref{eq:logFreqlog1plusz}}. 
The frequency interval at each $k$ is different.  
It scales by $(1+z_{k})$ with respect to that at $z_0=0$, 
 with a coarser frequency interval at a higher $z_k$. 
The assignment of the frequency and the redshift 
  to the computational grids 
  are illustrated in \Fig{\ref{fig:C21LRTRectGridRayTrace}}.  
The constant integer ratio 
  $[\Delta_{\log{(1+z)}}/\Delta_{\log{\nu}}] = C_1/C_2= S$ 
  governs 
  a constant shift in the $j$ index 
  of where $\nu_{21{\rm cm}}$ lies 
  (denoted by ${\rm ind}_{21{\rm cm}}$) 
  at each $z_k$.  
If at $k=0$, ${\rm ind}_{21{\rm cm}} =0$ 
  by a choice (\ie $\nu(0,0) = \nu_{21{\rm cm}}$),  
  then, for $k \geq 1$, ${\rm ind}_{21{\rm cm}}|_{z_{k}} = -kS$. 
Local 21-cm emission and absorption 
  at all redshifts can, therefore, be tracked. 

The algorithm can be optimised 
  to increase computational efficiency. 
For instance, 
 in certain post-reionisation epochs, 
 where foreground effects are insignificant  
 (\ie absence of significant foreground absorption, 
 emission, and line-continuum interaction), 
 the radiative transfer of the 21-cm line  
 can be performed simply by passing 
 on the invariant specific intensity 
 along the same $j$ index as the index $k$ descends to zero 
 in the computational lattice. 
The local comoving specific intensity in the observer frame 
  is calculated directly from invariant specific intensity. 

The computational efficiency 
  can further be boosted 
  by an OpenMP parallelisation of the C21LRT code,  
  when evaluating the frequency range (over the index $j$) 
  at each redshift (at a given $k$).   
Consistent results are obtained 
 using the OpenMP parallelised code 
  as those obtained by the serial execution 
  in all code verification tests. 

\subsection{Resolutions and convergence}
\label{subsec:reso} 
The resolutions used in our code are constrained by two main aspects, which are to resolve the 21-cm line profile locally and preserve the line feature during radiative transfer from higher to lower redshift. 

The frequency grid is set to resolve the local 21-cm profile at all redshifts where $n_{\rm HI}$ is positive. The redshift grid is set to ensure that there is enough overlap between the blue wing (higher frequency end) of the line profile in a given redshift cell ($z=z_2$) and the red wing (lower frequency end) of the next lower redshift cell ($z=z_1<z_2$). The resolutions we adopted for the calculations in this paper are listed in Table~\ref{tab:C21LRT_reso}. 

As explained in Sec \ref{subsubsec:LineProfile}, the line profiles determined with $v_{\rm turb}$ in this paper only act as a redistribution of the emission or absorption over a frequency scale which is much smaller than the frequency range of the presented spectra. The 21-cm spectra calculated with various line widths are expected to be almost identical. We therefore present 
the relative differences for $I_{\rm L}-I_{\rm C}$ when adopting various turbulent velocities in Table~\ref{tab:C21LRT_diff_I_L_I_C}. We note that the absolute value of $I_{\rm L}-I_{\rm C}$ can be very close to 0 at some frequencies (redshifts) ranges, including (i) at the beginning $z=z_{\max}$ and the end $z=z_{\min}$ of the radiative transfer calculation before conversion to the present-day values seen by an observer at $z=0$, and (ii) at $z\approx 11.04$ where the 21-cm signal changes from absorption to emission. When calculating the relative difference, we remove these frequencies (redshifts) ranges where the signal is too close to 0. In these ranges, the relative differences are about 0.5 to 1.3. 

Although the 21-cm spectra is usually presented in $I_{\rm L}-I_{\rm C}$-frequency diagrams. The quantity that is calculated and propagated is $I_{\rm L}$. We therefore also show the difference in $I_{\rm L}$ when adopting various turbulent velocities are listed in Table~\ref{tab:C21LRT_diff_I_L}. 

\begin{table}
\caption{The frequency and redshifts  resolutions adopted for the calculations in this paper. The frequency resolution is set to resolve the 21-cm line profile in frequency space at all relevant redshift points. The redshfit resolution is set to  preserve 21-cm spectral features during radiative transfer from the highest relevant redshift $z=z_{\max}$ to $z=0$.}
\begin{tabular}{lllll}
\hline
 $v_{\rm turb}$(${\rm km\,s^{-1}}$)  & 1000     & 100&  10 &  1 \\ 
\hline
   $\Delta_{\log(1+z)}$
   &$1\times10^{-4}$  &$1\times10^{-4}$ &$1\times10^{-5}$  &$1\times10^{-6}$\\
   $\Delta_{\log{\nu}}$ &$1\times10^{-5}$ &$1\times10^{-5}$ &$1\times10^{-5}$ &$1\times10^{-6}$\\
   \hline
\end{tabular}
\label{tab:C21LRT_reso}
\end{table}
\begin{table}
\caption{Relative difference in 21-cm line intensity ($I=I_{\rm L}-I_{\rm c}$) between scenarios with different turbulent velocity $v_{\rm turb}$. The relative difference is defined as, e.g. $\left|I (v_{\rm turb}=100~{\rm km\,s^{-1}})/I (v_{\rm turb}=1000~{\rm km\,s^{-1}})-1\right|$ averaged over the frequency range presented in \Fig\ref{fig:21_spectra_z0_various_v_turb} as determined by $z_{\max}$ and $z_{\min}$. 
}
\begin{tabular}{p{16mm}p{19mm}p{19mm}p{19mm}}
\hline
 Relative difference of\newline  ($I=I_{\rm L}-I_{\rm C}$)&
 $v_{\rm turb}$=1000 ${\rm km\,s^{-1}}$    &  $v_{\rm turb}$=100 ${\rm km\,s^{-1}}$ &  $v_{\rm turb}$=10 ${\rm km\,s^{-1}}$  \\ 
\hline
$v_{\rm turb}$=100 ${\rm km\,s^{-1}}$& $2.78\times10^{-4}$ & - &- \\
$v_{\rm turb}$=10 ${\rm km\,s^{-1}}$&$3.46\times10^{-4}$ & $1.56\times10^{-4}$ &- \\
$v_{\rm turb}$=1 ${\rm km\,s^{-1}}$&$3.57\times10^{-4}$ &$1.71\times10^{-4}$ &$1.55\times10^{-5}$\\
\hline
\end{tabular}
\label{tab:C21LRT_diff_I_L_I_C}
\end{table}
\begin{table}
\caption{Relative difference in 21-cm line intensity ($I=I_{\rm L}$) between scenarios with different turbulent velocity $v_{\rm turb}$. The relative difference is defined as, e.g. $\left|I (v_{\rm turb}=100~{\rm km~s}^{-1})/I (v_{\rm turb}=1000~{\rm km~s}^{-1})-1\right|$ averaged over the frequency range presented in \Fig\ref{fig:21_spectra_z0_various_v_turb} as determined by $z_{\max}$ and $z_{\min}$.
}
\begin{tabular}{p{16mm}p{19mm}p{19mm}p{19mm}}
\hline
 Relative difference of\newline  ($I=I_L$)&  $v_{\rm turb}$=1000 ${\rm km\,s^{-1}}$    
 &  $v_{\rm turb}$=100 ${\rm km\,s^{-1}}$ &  $v_{\rm turb}$=10 ${\rm km\,s^{-1}}$ \\ 
\hline
$v_{\rm turb}$=100 ${\rm km\,s^{-1}}$& $1.88\times10^{-6}$& - &- \\
$v_{\rm turb}$=10 ${\rm km\,s^{-1}}$&$2.49\times10^{-6}$ & $1.11\times10^{-6}$ &- \\
$v_{\rm turb}$=1 ${\rm km\,s^{-1}}$&$2.58\times10^{-6}$&$1.22\times10^{-6}$&$1.11\times10^{-7}$\\
\hline
\end{tabular}
\label{tab:C21LRT_diff_I_L}
\end{table}


\section{Code verification}
\label{app:C21LRTCodeVeri} 

\subsection{Continuum radiative transfer}
\label{subsec:codeveri-one-CMB} 
This test 
  is to verify that the effect of cosmological expansion 
  is properly account for in the absence of emission and absorption. 
The CMB has a blackbody spectrum, 
  described by a Planck function with a single parameter, a thermal temperature. 
Its properties is well established observationally, 
 and so it is chosen as a our test continuum. 

\subsubsection{Set-up}
\label{subsubsec:codeveri-one-CMB-setup}

A ray is traced from $z_{\rm emi} = 35.0$ to $z=0$, 
  with an initial specific intensity   
  $I_{\nu}|_{z_{\rm emi}} 
    = B_{\nu}({T_{\rm CMB}}\big|_{z_{\rm emi}})$.   
Without absorption and emission, 
  the spectral evolution of the CMB  
  is determined by the cosmological expansion only  
  and can be parametrised by the redshift $z$. 
The CMB temperature scales with $z$ as  
  $T_{\rm CMB}(z)/(1+z) = T_{\rm CMB, 0}$ in the FLRW universe, 
  where $T_{\rm CMB, 0} = 2.73~{\rm K}$ 
  at the current epoch \citep[][]{Spergel2003ApJS, Planck13}.  

\subsubsection{Results}
\label{subsubsec:codeveri-one-CMB-results}

\begin{figure}
\centering
\includegraphics[width=1.0\columnwidth]{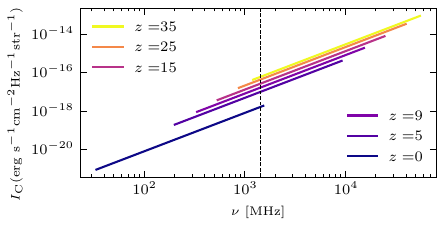}    
    \caption[C21LRT test results: cosmological transfer of the CMB spectrum]{The transfer of the CMB 
        from high to redshifts  
        (in log-log representation).   
    The frequency 
     of hyperfine 21-cm transition of \HI 
      in the local rest frame, 
      is marked \textcolor{red}{with the vertical black dashed line}.
          }  
\label{fig:CMB_appendix}
\end{figure}

\Fig~\ref{fig:CMB_appendix}
  shows the resulting CMB spectra at different cosmological epochs. 
The discrepancy between the computed and observed CMB spectra at $z=0$ 
  is smaller than 1 part in $10^{14}$ (i.e., comparable 
  with machine floating-point precision). 

\begin{figure}
\centering
\includegraphics[width=1.0\columnwidth]{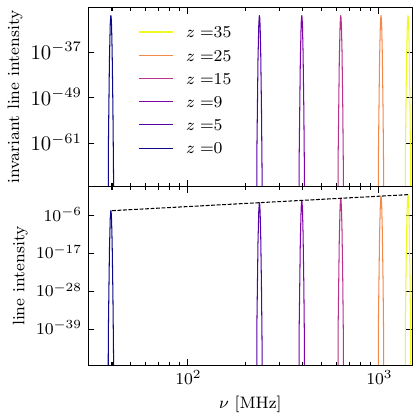}    
    \caption[C21LRT code verification test II results: 
        variations of the line properties at different redshifts]{
      The computed invariant line intensity 
      ${\mathcal I}_{\nu}$ (first panel, in units of ${\rm erg}\,{\rm s}^{-1}\,{\rm cm}^{-2}\,{\rm Hz}^{-4}\,{\rm str}^{-1}$)  
         and line intensity $I_{\nu}$ (second panel, in units of ${\rm erg}\,{\rm s}^{-1}\,{\rm cm}^{-2}\,{\rm Hz}^{-1}\,{\rm str}^{-1}$) 
         against $\nu$ in log-log scales, 
         in an expanding universe 
          without a line-of-sight medium 
         and an external radiation field. 
      The results show that ${\mathcal I}_{\nu}$ remains constant over redshift $z$,  $I_{\nu}$ scales with $\left[(1+z)/(1+z_{\rm emi})\right]^{3}$, and
        the line shape is preserved in the $\log{\nu}$ space, all 
        agreeing with the predictions of the analytical calculations. 
     }   
\label{fig:line_appendix}
\end{figure} 

\subsection{Line radiative transfer} 
\label{subsec:codeveri-two-lineprofile} 

This test is to verify 
  correct frequency shifting, broadening and 
  Lorentz-invariance-induced suppression of intensity 
  when the line is covariantly transported across the redshift space,  
  in the absence of absorption and emission,  
  via ray-tracing.   
  
\subsubsection{Set-up}
\label{subsubsec:codeveri-two-line-setup}

The line propagates down 
  in the redshift space from $z_{\rm emi}= 35.0$ to $z=0$.   
The frequency shift of the line as it propagates   
   is governed by a $\left[(1+z)/(1+z_{\rm emi})\right]$ factor 
    (cf. \eqn{\ref{eq:cosmologicalFreqShift}}), 
    and this frequency shift  has several consequences. 
(i) Without absorption and emission, 
  $\mathcal{I_{\nu}}(=I_\nu/\nu^3)$ of the line is invariant 
  in a convariant radiative transfer.    
The comoving specific intensity of the line ${I_{\nu}}$ 
   will therefore decrease accordingly, 
   following $\left[(1+z)/(1+z_{\rm emi})\right]^{3}$,   
(ii) The line width, expressed as a frequency spread, 
  will be squeezed\footnote{
Cosmological redshift $z$ is defined as 
\begin{align}  
 (1+z_{\rm emi}) & \equiv  \frac{\lambda_{\rm obs}}{\lambda_{\rm emi}} 
  = \frac{\nu_{\rm emi}}{\nu_{\rm obs}} \nonumber \ ,  
\end{align} 
  with a rest-frame 
  observer at 
  $z_{\rm obs} = 0$. 
This gives  
\begin{align} 
   (1+z_{\rm emi}) &  
   = \frac{\nu_{\rm emi}+\delta \nu_{\rm emi}}{\nu_{\rm obs}+\delta\nu_{\rm obs}} 
   = \frac{\nu_{\rm emi}}{\nu_{\rm obs}}
   \left[
   \frac{1+(\delta \nu_{\rm emi}/\nu_{\rm emi})}{1+(\delta \nu_{\rm obs}/\nu_{\rm obs}))} 
   \right]  \nonumber \ ,  
\end{align} 
 where $\delta \nu_{\rm emi}$ is a frequency displacement 
   from $\nu_{\rm emi}$,   
  and $\delta \nu_{\rm obs}$ is the corresponding frequency displacement 
   from $\nu_{\rm obs}$ measured by the observer. 
It follows from the two expressions that 
\begin{align}  
   \frac{\delta \nu_{\rm emi}}{\nu_{\rm emi}}
  &  = \frac{\delta \nu_{\rm obs}}{\nu_{\rm obs}} \ ,  \nonumber 
\end{align} 
 which implies that 
\begin{align}  
    \frac{\delta \nu_{\rm obs}}{\delta \nu_{\rm emi}} 
     &  = \frac{\nu_{\rm obs}}{\nu_{\rm emi}} = \frac{1}{(1+z_{\rm emi})} \ .  \nonumber 
\end{align}
If $\delta \nu_{\rm emi}$ is the marker of the width of a line, 
  centred at $\nu_{\rm emi}$, 
  emitted from $z_{\rm emi}$, 
  the width of the line will reduce when measured by the observer 
  in its local reference frame  
  (for $z_{\rm emi} > z_{\rm obs} = 0$ in an expanding Universe).    
} 
  (as opposed to the stretch when the line width 
   is expressed as a wavelength spread) 
   by a factor of  
   $\left[(1+z)/(1+z_{\rm emi})\right]$. 
The line width when expressed as velocity spread, 
  however, remains unchanged.  
(iii) The line shape is distorted 
  when it is expressed in terms of $\nu$,   
  but preserves when it is expressed in terms of $\log{\nu}$.  
  
We considered a line, which has a guassian profile 
  initially at $z_{\rm emi}=35.0$ 
  and is centred at $\nu = 1.42~{\rm GHz} = \nu_{21{\rm cm}}$. 
It has a width $\Delta {\nu_{\rm D}} = 4.738~{\rm MHz}$, 
  corresponding to a velocity dispersion\footnote{ 
The velocity dispersion of galaxies inside massive galaxy clusters  
  \citep[e.g. the Coma cluster][]{Struble1999GCvelodisp} 
  could have velocity dispersion 
  $\Delta v \approx 1000~{\rm km}\,{\rm s}^{-1}$.  
A large value of $\Delta v (=1000 ~{\rm km}\,{\rm s}^{-1})$ 
  is therefore selected to verify 
  that the code has the dynamical range to handle 
  this extreme level of line broadening.    
The velocity dispersion caused 
  by the differential motions 
  within a galaxy is smaller, generally 
  in the range $\Delta v \sim 100 - 400~{\rm km}\,{\rm s}^{-1}$ 
  \citep[see \eg][]{Bezanson2012GalaxiesVeloDisp}.  
This gives rise to a frequency spread of    
  $\Delta {\nu_{\rm D}} = 0.474 - 1.89~{\rm MHz}$. 
} 
of $\Delta v =1000~{\rm km}\,{\rm s}^{-1}$. 
The initial value of the peak specific intensity of the line 
  is set to be 
   $I_\nu = 1.0~{\rm erg}\,{\rm s}^{-1}\,{\rm cm}^{-2}\,{\rm Hz}^{-1}\,{\rm str}^{-1}$  
   (in the local rest frame) at $z_{\rm emi}=35.0$.

\subsubsection{Results}
\label{subsubsec:codeveri-two-line-results}

\begin{figure}
\centering
\includegraphics[width=1.0\columnwidth]{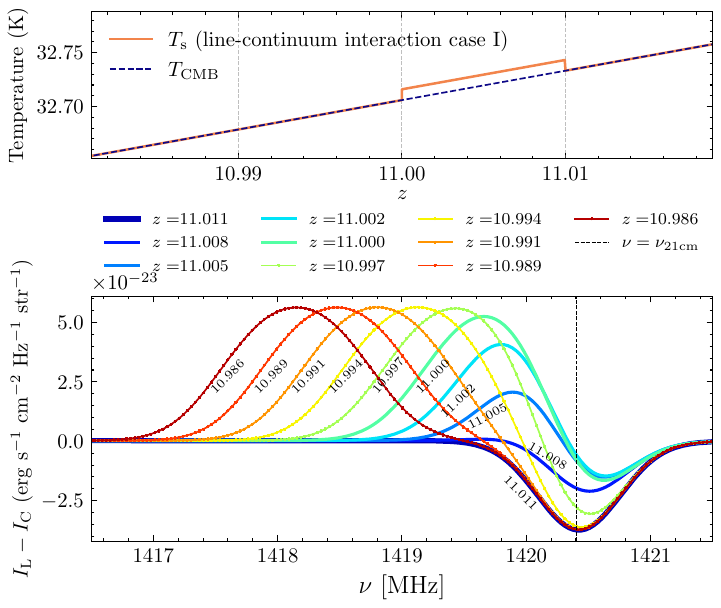}
\caption[caseI]
{Line continuum interaction case I. The top panel shows the input spin temperature for our line continuum interaction calculation. The \HI density is still the same as in \Fig\ref{fig:input_21CMFAST}. The spin temperature $T_{\rm s}$ is set to be 0.01 K higher than $T_{\rm CMB}$ between $z=11.00$ and $z=11.01$, and $T_{\rm s}=T_{\rm CMB}$ at $z>11.01$ or $z<11.00$. The turbulent velocity $v_{\rm turb}$ is set to be 100 ${\rm km\,s^{-1}}$. This would generate a narrow emission signal in the local rest frame. The bottom panel shows the spectra calculated with C21LRT in the local rest frame of over redshift $z=10.986-11.011$. See main text for the how line continuum interaction shaped these spectra.} 
\label{fig:caseI} 
\end{figure}

\Fig~\ref{fig:line_appendix}
  shows the profiles of the line, 
  in invariant specific intensity $\mathcal{I}_\nu$ (in units of ${\rm erg}\,{\rm s}^{-1}\,{\rm cm}^{-2}\,{\rm Hz}^{-4}\,{\rm str}^{-1}$)
  and comoving specific intensity $I_\nu$ (in units of ${\rm erg}\,{\rm s}^{-1}\,{\rm cm}^{-2}\,{\rm Hz}^{-1}\,{\rm str}^{-1}$), 
  at selected redshifts. 
As shown, the invariant specific intensity 
  remains constant (in the top panel) but 
  the specific intensity decreases 
  (with a trend indicated by the dashed straight line in the bottom panel) 
  as the line  propagates from the high to low redshifts. 
The shape of the line is preserved in the $\log \nu$ representation. 
These are the same as expected from theoretical consideration. The residuals of $I_{\nu}$ and $\nu$, 
  which are calculated 
  by subtracting the ratio of 
  the computed values 
  to their corresponding analytical values 
  by unity, 
  attain a level below $10^{-14}$, 
  reaching machine floating-point precision.  
The variations 
  of the line frequency, 
  characterised by the full-width-half-maxima,  
  ${\rm FWHM}_{\nu}$ in frequency and 
  ${\rm FWHM}_{v}$ in velocity, 
  also agree with the analytical values 
  at a level of $10^{-13}$.

\section{Line-continuum interaction}
\label{app:LineContinuumDiscussion}
\begin{figure}
\centering
\includegraphics[width=1.0\columnwidth]{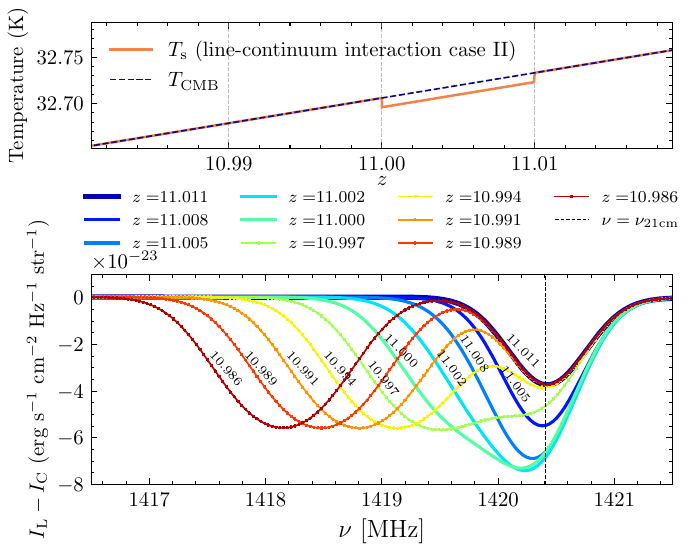}
\caption[caseII]
{Line continuum interaction case II. Similar to case I in \Fig~\ref{fig:caseI}, the top panel shows the input spin temperature for the second case of line continuum interaction calculation. The \HI density is still the same as in \Fig\ref{fig:input_21CMFAST}. The spin temperature $T_{\rm s}$ is set to be 0.01 K lower than $T_{\rm CMB}$ between $z=11.00$ and $z=11.01$, and $T_{\rm s}=T_{\rm CMB}$ at $z>11.01$ or $z<11.00$. The turbulent velocity $v_{\rm turb}$ is set to be 100 ${\rm km\,s^{-1}}$. This would generate a narrow absorption signal in the local rest frame. The bottom panel shows the spectra calculated with C21LRT in the local rest frame of over redshift $z=10.986-11.011$. See main text for the how line continuum interaction shaped these spectra.}
\label{fig:caseII} 
\end{figure}


Using the optical depth parametrisation  
also ignores the convolution of the line transfer and the continuum transfer 
\citep[see][]{Tucker1977Book,Hummer1985ApJ,Wu2001MNRAS}, 
which are characterised by different opacities. For radiative transfer through structures with a non-uniform 
line-of-sight velocity field,  
the convolution of the line transfer and the continuum transfer 
can give rise to complex spectra
\citep[see][]{Mihalas1978book,Rybicki1978ApJ,Hummer1985ApJ}. 
The line-continuum interaction also convolves with cosmological evolution (the time and frequency ordering mentioned in Sec \ref{subsec:optical_depth_Discussion}). To illustrate how all of these effect are naturally taken into consideration within C21LRT calculations, we present two cases of 21-cm line and CMB continuum interaction, one exhibiting an emission line signal and one with absorption line signal.

To create an emission 21-cm signal, we set the spin temperature as $T_{\rm s}=T_{\rm CMB}+0.01$~K in a narrow redshfit range (from $z=11.00$ to $z=11.01$), and $T_{\rm s}=T_{\rm CMB}$ everywhere else, as shown in the top panel of \Fig\ref{fig:caseI}. The \HI density is still the same as the default model (calculated from the ionisation fraction $x_{\rm i}$). We adopt a turbulent velocity of $v_{\rm turb}=~100 {\rm km\,s^{-1}}$, corresponding to FWHM$_{\nu}=0.7889$ MHz in the local rest frame.
The 21-cm spectra ($I_{\rm L}-I_{\rm C}$) in the rest frame of $z=11.011,\ 11.008,\ ......\ 10.986$, saved when the radiative transfer calculation has been carried out to these redshfit points are presented in the bottom panel. These spectra are color-coded as well as annotated, with the thick bluer lines for higher redshift starting with $z=11.011$ and the thin red line with dot marker for lower redshift ending at $z=10.986$. For all the presented spectra, there is a Gaussian function shaped dip at $\nu_{\rm 21cm}$. This is because the radiative transfer has only been carried out to this redshift (this is the same as the sharp decrease at $\nu_{\rm 21cm}$ in \Fig\ref{fig:21_spectra_development}). The spectrum at $z=11.011$ essentially has no 21-cm signal except for the dip. The 21-cm signal then gradually rises near $\nu_{\rm 21cm}$ first at $z=11.008$ to  $z=11.000$. These features would be lost if line broadening and radiative transfer effects are not account for. The emission bump then gradually moves towards lower frequency from $z=11.000$ to $z=10. 986$ and the dip gradually recovered to the original shape.

Similarly, we set the spin temperature as $T_{\rm s}=T_{\rm CMB}-0.01$~K from $z=11.00$ to $z=11.01$ to create an absorption 21-cm signal, as shown in the top panel of \Fig\ref{fig:caseII}. The 21-cm spectrum also start with a dip centred at $\nu_{\rm 21cm}$ at $z=11.011$ (blue thick line in the bottom layer of the bottom panel). Then true absorption feature first develops near $\nu_{\rm 21cm}$ at $z=11.008$  and gradually widens. The true absorption feature and the original dip due to calculation creates a valley like shape at $z=10.997$ (green line with dot marker). Then the absorption feature fully develops and moves towards lower redshift.

For both cases, there was no significant true features at $z>11.011$. 
The absorption and emission features will continue to move towards lower frequency as radiation propagate towards $z<10.986$. We therefore only present the spectra with $10.986<z<11.011$. The frequency range in the spectra where the convolution effect between line broadening and radiative transfer with continuum emission is the most prominent is jointly determined by the line profile, $T_{\rm s}-T_{\rm CMB}$, and the redshift range where $T_{\rm s}-T_{\rm CMB}$ is non-zero. We intentionally adjusted these factors so that the spectra in \Fig\ref{fig:caseI} and \Fig\ref{fig:caseII} are clear and straightforward to interpret. The line continuum interaction effects are still present if we adopt a more complicated $T_{\rm s}$ and $x_{\rm i}$ model, albeit less straightforward for interpretation.


\bsp	
\label{lastpage}
\end{document}